\documentclass[10pt]{iopart}
\usepackage{iopams}
\usepackage{graphicx}
\usepackage{cite}

\def\vep{\varepsilon}

\def\vf{\varphi}

\def\be{\begin{equation}}
\def\ee{\end{equation}}
\def\beq{\begin{equation}}
\def\bea{\begin{eqnarray}}
\def\eea{\end{eqnarray}}
\newcommand{\beqn}{\begin{eqnarray}}
\newcommand{\eeqn}{\end{eqnarray}}
\newcommand{\beqnn}{\begin{eqnarray*}}
\newcommand{\eeqnn}{\end{eqnarray*}} 

\begin{document}
 
\title{Energy--time and frequency--time uncertainty relations: exact inequalities}

\author{V V Dodonov and A V Dodonov}

\address{Instituto de F\'{\i}sica, Universidade de Bras\'{\i}lia, PO Box 04455,
70910-900, Bras\'{\i}lia, DF, Brazil}
\ead{vdodonov@fis.unb.br} 

\begin{abstract}
We give a short review of known exact inequalities that can be interpreted as `energy--time'
and `frequency--time' uncertainty relations. In particular we discuss a precise form of signals
minimizing the physical frequency--time uncertainty product. Also, we calculate the
`stationarity time' for mixed Gaussian states of a quantum harmonic oscillator, showing explicitly that pure quantum states are
`more fragile' than mixed ones with the same value of the energy dispersion.
The problems of quantum evolution speed limits, time operators and measurements of energy and time are briefly discussed, too.

\end{abstract}

{\em This is our present to Margarita Alexandrovna and Vladimir Ivanovich Man'ko
on occasion of their 75th birthdays.}



\section{Introduction}

The energy--time uncertainty relation (ETUR)
\be
\Delta E \Delta T  \gtrsim h
\label{ETh}
\ee
is one of the most famous and at the same time most controversial formulas of quantum theory.
It was introduced by Heisenberg \cite{Heis27} together with his coordinate--momentum
uncertainty relation 
\footnote{Heisenberg and Bohr used the symbols $=$ or $\sim$ instead of $\gtrsim$.}
\be
\Delta x \Delta p  \gtrsim h.
\label{xph}
\ee
The importance of both relations for the interpretation of quantum mechanics was emphasized by Bohr \cite{Bohr28}.
However, the further destiny of relations (\ref{ETh}) and (\ref{xph}) turned out quite different.
A strict formulation of relation (\ref{xph}) was found almost immediately by Kennard \cite{Kenn27} and Weyl \cite{Weyl28}
in the form of inequality
\be
\Delta x \Delta p \ge \hbar/2,
\label{xph2}
\ee
where $\Delta x$ and $\Delta p$ are well-defined quantities, namely mean-square deviations (dispersions). 
A liitle later, more general
inequalities were derived by Robertson \cite{Rob-ABC,Rob-un} and Schr\"odinger \cite{Schr-un}.
Actually, the meaning of inequality (\ref{xph2}) is rather different from the original
thoughts of Heisenberg and Bohr related to {\em measurements\/} of canonical variables
(see e.g. \cite{Helstrom76,Holevo-book,Ozawa04,Busch14}). Nonetheless (\ref{xph2}) is
a well established consequence of the quantum mechanical formalism.

On the contrary, the physical and mathematical meanings of inequality (\ref{ETh}) appeared to be much less clear than that
of (\ref{xph2}) (and less clear than Heisenberg, Bohr and other creators of quantum mechanics thought initially). 
The main reason is that, in fact, there are several quite different physical problems where relations like
(\ref{ETh}) can arise, and in each concrete case the meaning of the quantities
standing on the left-hand side proves to be different. This was clearly demonstrated for the first time 
 by Mandelstam 
\footnote{We use the name Mandelstam as it appears in western literature;
in Russian literature it is written as Mandelshtam.}
and Tamm \cite{TM} 
and by Fock and Krylov \cite{FoKry}, and many authors arrived at the same conclusions later
\cite{AharBohm61,Allcock,Bauer,Busch90-1}.

The aim of this paper is to give a concise review of some results, where the strict inequality
sign ($\ge$) replaces the symbol $\gtrsim$ of approximate (by an order of magnitude) inequality in relation (\ref{ETh}).
Such a mini-review seems to suit well for the inclusion to the special issue celebrating 75+75 years of Margarita and 
Vladimir Man'ko, because they have been always interested in the related subjects: see, e.g., references
\cite{blackbook,183,MAM06} where different aspects of time--energy and time--frequency uncertainty 
relations were discussed (together with many other topics).
Our initial goal was to extend or generalize some results obtained many years ago in \cite{blackbook,183}. These generalizations
are given in sections \ref{sec-decay2}-\ref{sec-Eb}.
However, it seems reasonable
to describe the most important achievements of many other authors as well, providing the relevant references.
Although many reviews on the ETUR were published during decades \cite{Price-book,MugaLeavens,Muga-book,Hilg05,Sen14},
we hope nonetheless  that this mini-review with the updated reference list will be also useful for many readers as the complementary
source of information.

\section{Mandelstam--Tamm inequalities and their application to the problem of decay} 
\label{sec-MT}

The first rigorous formulations of relation (\ref{ETh}) was given by Mandelstam and Tamm \cite{TM}. 
They used the uncertainty relation for two arbitrary Hermitian operators $\hat{A}$ and $\hat{B}$ 
(derived for the first time by Robertson \cite{Rob-ABC})
\be
\Delta A \Delta B \ge \frac12\left|\langle [\hat{A},\hat{B}]\rangle\right|
\label {(3)}
\ee
Then, choosing $\hat{B}=\hat{H}$ (where $\hat{H}$ is the system
Hamiltonian) and remembering that operator $(i/\hbar)[\hat{H}, \hat{A}]$ is the operator of
the rate of change of the quantity $A$, i.e.,  $(i/\hbar)[\hat{H}, \hat{A}]=d\hat{ A}/dt$  (provided operator $\hat{A}$
does not depend on time explicitly), one can rewrite (\ref{(3)}) as 
\be
\Delta E \Delta t_A \ge \hbar/2,
\label{(299)}
\ee
where $\Delta E \equiv \Delta H$ and
\be
 \Delta t_A \equiv \Delta A/|\langle d\hat{ A}/dt\rangle|.
\ee
The meaning of inequality (\ref{(299)}) is that it yields an estimate of the time interval 
required for a significant change in the average value of observable $A$:
by an amount of the order of the mean squared variation. 
It results in the statement that  `a dynamical quantity cannot change,
remaining always dispersionless' \cite{TM}.

Relation (\ref{(299)}) may seem a little artificial at first glance, since operator $\hat{A}$ may be quite arbitrary.
However, there exists at least one important specific choice of this operator. Namely, let us consider,
following \cite{TM}, the projector $\hat{L}$ on some initial quantum state $|\psi(0)\rangle$:
 $\hat{L} = |\psi(0)\rangle\langle \psi(0)|$. 
Since $\hat{L}$ is the projection operator, $\hat{L}=\hat{L}^2$ and
\[
\Delta L \equiv \sqrt{\overline{(L^2)} -(\overline{L})^2} =
\sqrt{\overline{L} -(\overline{L})^2}.
\]
Then inequality (\ref{(299)}) assumes the form
\be
-\frac{\hbar}{2}\frac{d\overline{L}}{dt} =\frac{\hbar}{2}\left|\frac{d\overline{L}}{dt}\right|
\le \Delta E \sqrt{\overline{L} -(\overline{L})^2}.
\label{eq-L}
\ee
Integrating (\ref{eq-L}) with account of the initial condition $\overline{L}(0)=1$,
Mandelstam and Tamm obtained relations
\begin{equation}
\Delta Et/\hbar\ge\pi /2-\arcsin\sqrt {Q(t)}, \quad 
0\le t<\infty ,
\label{330a}
\end{equation}
\begin{equation}
Q(t)\ge\cos^2(\Delta Et/\hbar ), \quad 0\le t\le\pi
\hbar /2\Delta E,
\label{330b}
\end{equation}
where
\be
Q(t) \equiv \overline{L}(t) \equiv \langle\psi(t)|\hat{L}|\psi(t)\rangle =
|\langle \psi(0)|\psi(t)\rangle |^2. 
\label{defQ}
\ee

One may interpret function $Q(t)$ as the probability to remain in the
initial state $|\psi(0)\rangle$. In this case it seems natural to define the half-decay period 
$T_{1/2}$ by means of the relation $Q\left(T_{1/2}\right)=1/2$.
 Then  (\ref{330b}) results in the inequality
\begin{equation}
T_{1/2}\Delta E\ge\pi\hbar /4\approx 0.785\hbar
\label{332}
\end{equation}
with well defined quantities $T_{1/2}$ and $\Delta E$.

An immediate important consequence of inequality  (\ref{330b})
is the impossibility of strictly exponential decay
\begin{equation}
Q(t)=\exp(-t/\tau )
\label{327}
\end{equation}
for realistic physical systems with a finite energy dispersion $\Delta E$ \cite{TM,FoKry,Fleming73,Fonda78}.
Indeed, it follows from (\ref{330b}) that the law (\ref{327})
can be realized only approximately and for sufficiently big values of time satisfying the inequality
\be
t> \tau\ln\left[1+ \hbar^2/(2\tau\Delta E)^2\right]. 
\label{t-tau}
\ee

Generalizations of relations (\ref{330a}) and (\ref{330b}) to the case of time-dependent Hamiltonians were obtained
in \cite{Pfeifer95}.
Noticing that inequality (\ref{(299)}) becomes meaningless in the case of $\Delta E=\infty$,
 Mandelstam and Tamm wrote in \cite{TM} that `it would be desirable to find a more general relation
of the same type as (\ref{(299)})'. In the following sections we demonstrate how this can be done
and what concrete generalizations were proposed.

\section{Decay laws and spectral distributions}

The relations between the decay law $Q(t)$ and the energy spectrum of the system
were established by Krylov and Fock \cite{FoKry}.
Suppose that we know the decomposition of vector $|\psi(0)\rangle$ over the
energy eigenstates:
\be
|\psi (0)\rangle =\int a(E)|E\rangle\,dE, \quad \langle E'|E\rangle =\delta(E-E').
\ee
Then 
\be
|\psi (t)\rangle =\int a(E)\exp\left(-iEt/\hbar\right)|E\rangle\,dE
\label{psitE}
\ee
and the nondecay amplitude $\chi(t)=\langle\psi(0)|\psi(t)\rangle$
(called `integrity amplitude' in \cite{Fleming73})
 can be expressed as the
Fourier transform of the positive energy distribution function $P(E)=|a(E)|^2$:
\begin{equation}
\chi (t) \equiv \langle\psi(0)|\psi(t)\rangle =\int P(E)\exp\left(-iEt/\hbar\right)\,dE.
\label{326}
\end{equation}
Its consequence is the identity
\be
\int P(E)dE=1
\label{normP}
\ee

The probability of finding the system in the initial state at time $t$ equals 
$Q(t)=|\chi (t)|^2$.
The energy variance can be calculated as
\be
(\Delta E)^2=\int\left(E-\bar E\right)^2P(E)\,dE
\label{varE}
\ee
with
\be
\bar {E}=\int EP(E)\,dE.
\label{minE}
\ee
 Fock and Krylov \cite{FoKry} proved that the
necessary and sufficient condition of decay (the property $Q(t)\to 0$ for $t\to\infty$)
is the continuity of the integral energy distribution function
$\tilde{P}(E)= \int^{E} P(\vep)d\vep$.
This means, in particular, that the energy spectrum must be continuous, in
order that function $P(E)$ would not contain terms like $\delta(E - E_0)$ corresponding
to discrete energy levels. 

It is known that the exponential decay law (\ref{327}) corresponds to the Lorentzian
energy distribution 
\begin{equation}
P(E)=\frac {\Gamma /2\pi}{\left(E-E_0\right)^2+\Gamma^2/4}.
\label{328}
\end{equation}
The peak width $\Gamma$, defined as the length of the energy interval
where $P(E) \ge P(0)/2$, is related to the lifetime $\tau$ by the equality
\be
\tau\Gamma = \hbar.
\label{(329)}
\ee
However, distribution (\ref{328}) is an idealization (although very good one in many practical cases), 
because it results in the relation $\Delta E=\infty$. Another drawback of distribution (\ref{328}) is
that it implies that the energy spectrum stretches from $-\infty$ to $\infty$
(only under this assumption integral (\ref{326}) yields the exponential function of time).
But the energy of real physical systems is limited from below, and this fact leads to violations of the exponential decay law
for $t\to\infty$, when $Q(t)\sim t^{-\beta}$ with some constant $\beta$ depending on the concrete form of 
the energy spectrum \cite{Khalfin} (see \cite{Fonda78,Newton61,Terentiev72,Peres80,Garcia96,Campo11} for later discussions and reviews).
Moreover, an oscillatory decay is also possible \cite{Peshkin14}.

A simple example of the `decay' that never has the exponential form was given by Bhattacharyya \cite{Bhat83}:
a freely expanding Gaussian wave packet in one dimension has the energy distribution function
\be
P(E) = \left(\sqrt2 \pi  E \Delta E\right)^{-1/2}\exp\left(-E/\sqrt{2}\Delta E\right)
\ee
and the nondecay probability
\be
Q(t) =\left(1 +2(\Delta E)^2 t^2/\hbar^2\right)^{-1/2}.
\ee
In this case $T_{1/2}\Delta E =\hbar\sqrt{3/2}$.

The remarkable formula (\ref{326}) has many interesting consequences. One of them was obtained by Luo \cite{Luo-jpa05}.
We reproduce it in a slightly modified form. Let us consider function $A(y;t)=\left(\mbox{Re}\left[\exp(iy)\chi(t)\right]\right)^2$, where
$y$ is an auxiliary variable. Obviously $Q(t)=|\chi(t)|^2$ is the maximal value of $A(y;t)$ for the fixed value of $t$. 
On the other hand formula (\ref{326}) yields (we put $\hbar=1$ here)
$A(y;t) = \left[ \int \cos(y-Et) P(E) dE \right]^2$.
Using the Schwartz inequality
\be
\int |f(x)|^2 dx \int |g(x)|^2 dx \ge \left|\int \mbox{Re}\left[f(x)g^*(x)\right] dx \right|^2
\label{(283)}
\ee
with $x=E$, $f(E)= \cos(y-Et) \sqrt{P(E)}$ and $g(E) =\sqrt{P(E)}$ we get, taking into account (\ref{normP}),
\beqnn
A(y;t) &\le&  \int \cos^2(y-Et) P(E) dE
 = \frac12  \int \left[1 + \cos(2y-2Et)\right] P(E) dE \\
& = & \frac12  \left(1 + \mbox{Re}\left[\exp(2iy)\chi(2t)\right] \right) \le \frac12 \left[1+|\chi(2t)|\right].
\eeqnn
Finally we have
\be
Q(t) \le \frac12\left[1 +\sqrt{Q(2t)}\right]
\label{Luo}
\ee
Some interesting consequencies of this inequality were discussed in \cite{Luo-jpa05}. 
The importance of inequality (\ref{Luo}) consists in the fact that
it forbids many decay laws that one could invent `from a head'. For example, it clearly
forbids the instantaneous decay, such that $Q(t)=1$ for $t<t_*$ and $Q(t)=0$ for $t>t_*$.
However, it is not the strongest possible inequality, because it 
does not forbid the exponential decay law  (\ref{327}), which is forbidden by (\ref{330b}) for short times.
On the other hand, looking at (\ref{330b}) one could suppose that the none-decay probability could behave
as $Q(t) \approx 1 - at^b$  with $b\ge 2$ for $t\to 0$. But inequality (\ref{Luo}) can be fulfilled in this limit
for $b\le 2$ only. So, only a parabolic time dependence of $Q(t)$ is permitted for short times.
This is clear from the well known short-time Taylor expansion 
\be
\fl
Q(t) = \left|\langle \psi_0|\exp\left(-\frac{i\hat{H}t}{\hbar}\right)|\psi_0\rangle\right|^2  
 \approx \left| 1 -\frac{it \langle\hat{H}\rangle}{\hbar}
-\frac{t^2\langle\hat{H}^2\rangle}{2\hbar^2} \right|^2  
=
1-\frac{t^2}{\hbar^2}\left(\langle\hat{H}^2\rangle -\langle\hat{H}\rangle^2\right)
\ee

\section{Wigner's approach}
\label{sec-Wigner}

Wigner \cite{Wig72} studied the function $\chi_u(t)=\langle u|\psi(t)\rangle$, i.e.,
 the probability amplitude of finding the quantum system described by the vector $|\psi(t)\rangle$
in some arbitrary fixed state $|u\rangle$.  The Fourier transform of function $\chi_u(t)$ was 
defined as (let us put here $\hbar=1$)
\be
\eta(E)= (2\pi)^{-1/2}\int \chi_u(t)\exp(iEt) dt. 
\label{(318)}
\ee
If the energy spectrum (i.e., the spectrum of the Hamiltonian operator) is absolutely continuous and bounded 
from below, then, taking $E_{min}=0$, one can expand vector $|u\rangle$  over eigenstates $|E\rangle$
of the Hamiltonian operator as follows,
\be
|u\rangle = \int_0^{\infty} b(E)|E\rangle dE.
\label{(319)}
\ee
Taking into account equation (\ref{psitE}) we have
\be
\eta(E)= (2\pi)^{1/2} b^*(E)a(E).
\label{(320)}
\ee
Wigner defined the characteristic time $\tau_W$ of the system's stay
in the state $|u\rangle$ and the energy dispersion $\vep_W$ as follows 
(assuming that the initial time is zero):
\be
\tau_W^2= \frac{\int_0^{\infty} t^2 |\chi_u(t)|^2 dt} {\int_0^{\infty} |\chi_u(t)|^2 dt},
\label{(321)}
\ee
\be
\vep_W^2= \frac{\int_0^{\infty} (E-E_0)^2|\eta(E)|^2 dE }{ \int_0^{\infty} |\eta(E)|^2 dE},
\label{(322)}
\ee
$E_0$ being an arbitrary parameter. If $\eta(0)\neq 0$, then function $\chi_u(t)$ behaves
as $1/t$ for $t \to\infty$ and $\tau=\infty$. Therefore Wigner considered 
 the states satisfying the restriction $\eta(0)=0$. He obtained the formula
\be
\tau_W^2 = \frac{\int_0^{\infty}|d\eta/dE|^2 dE }{ \int_0^{\infty} |\eta(E)|^2 dE}.
\label{(323)}
\ee
Using the Schwartz inequality (\ref{(283)})
he arrived at the inequality (recovering the Planck constant)
\be
\vep_W\tau_W > \hbar/2.
\label{(324)}
\ee
The equality in (\ref{(324)}) cannot be achieved due to the boundedness
 of the energy spectrum and the restriction $\eta(0)=0$. 
(The equality $\vep_W\tau_W = \hbar/2$ holds for the Gaussian distributions; 
but these distributions obviously do not satisfy the imposed restrictions.)
The minimal possible value of the product $\vep_W\tau_W$ depends on the parameter $E_0$.
For $E_0=0$ this minimal value equals \cite{Wig72}
$(\vep_W\tau_W)_{min}^{E_0=0}=3\hbar/2$. 
It is achieved for the function $\eta_0(E) = E\exp[-3E^2/(4\vep^2)]$.

A comparison of the M--T and Wigner UR was performed
 in \cite{Bauer} for free one-dimensional wave packets.
 It was shown that both methods lead to almost identical results. 
For the recent analysis of the Mandelstam--Tamm UR one can consult \cite{Gray05}.

\section{Different decay times of unstable systems}
\label{sec-decay2}

Besides the half-decay time $T_{1/2}$ and the Wigner time $\tau_W$ (\ref{(321)}), 
many other definitions of the decay time are possible. 
Fleming \cite{Fleming73} suggested to use the quantity
\begin{equation}
\tau_0=\int_0^{\infty}Q(t)\,dt
\label{333}
\end{equation}
where the non-decay probability $Q(t)$ was defined by equation (\ref{defQ}).
Definition (\ref{333}) gives the lifetime $\tau_0=\tau$ for the exponential function (\ref{327}). 
Taking into account inequality (\ref{330b}) and writing $z= {\pi\hbar }/(2\Delta E)$
 we obtain the following lower bound  for $\tau_0$ \cite{Fleming73}:
\begin{equation}
\tau_0\ge\int_0^{z}Q(t)\,dt\ge\int_
0^{z}\cos^2\left({\Delta Et}/{\hbar}\right)\,dt=\frac {\pi\hbar}{4\Delta E}.
\label{334}
\end{equation}
A weaker inequality $\tau_0 \Delta E \ge \hbar/2$
was obtained in \cite{Grabow84}. 

The most strong inequality with an achievable lower bound was given in study \cite{Gisl85}.
 First we note that the consequence of equations (\ref{defQ}), (\ref{326}) and  (\ref{333}) is the
formula
\be
\tau_0 =\pi\hbar \int_{-\infty}^{\infty}dE [P(E)]^2.
\label{(336)}
\ee
Therefore one has to minimize the functional
\[
I (P) =  \int_{-\infty}^{\infty}dE [P(E)]^2
\left[ \int_{-\infty}^{\infty}dE P (E) E^2\right]^{1/2},
\]
where the origin of the energy scale is chosen in such a way that $\overline{E} = 0$.
The value of functional $I(P)$ is not changed if instead of function $P(E)$ we
use the function $P_{\lambda}(E) = \lambda P(\lambda E)$ with an arbitrary (positive) parameter $\lambda$. 
Choosing this parameter in such a way that $\langle E^2\rangle =1$, 
 we arrive at the problem of minimizing the functional 
\be
\tilde{I}(P) = \int_{-\infty}^{\infty}dE [P(E)]^2
\label{(337)}
\ee
under the following auxiliary conditions (here we use dimensionless variables and $\hbar=1$):
\be
\int_{-\infty}^{\infty}dE P(E) = \int_{-\infty}^{\infty}dE P(E) E^2 =1,
\label{(338a)}
\ee
\be
\int_{-\infty}^{\infty}dE P(E) E =0,  \quad P(E) \ge 0.
\label{(338b)}
\ee
It was shown in \cite{Gisl85} that the extremal function is the truncated parabola
\be
P_0(E) = \left\{ 
\begin{array}{cc}
(\sqrt{45}/20)\left(1-E^2/5\right), & |E| \le \sqrt{5}
\\
0, & |E| > \sqrt{5}
\end{array}
\right. .
\label{(339)}
\ee
To prove this result, let us consider some function $R(E)$ satisfying conditions (\ref{(338a)}) and (\ref{(338b)}).
If $f(E)= R(E) -P_0(E)$, then
\be
\tilde{I}(R) =  \tilde{I}(P_0 +f) =  \tilde{I}(P_0) +  \tilde{I}(f)
+2 \int_{-\infty}^{\infty}dE P_0(E) f(E).
\label{(340)}
\ee
It is obvious from (\ref{(337)}) that $ \tilde{I}(f) \ge 0$; moreover, $ \tilde{I}(f) >0$ if the function $f(E)$
can turn into zero only for a set of zero measure. Let $g(E)$ be the non-truncated parabola (\ref{(339)}).
 Since $R(E)$ and $P_0(E)$ satisfy conditions (\ref{(338a)}) and (\ref{(338b)}),
\be
\int_{-\infty}^{\infty}dE f(E) E^n =0, \quad n=0,1,2.
\label{(341)}
\ee
Consequently,
\be
\fl
0 = \int_{-\infty}^{\infty}dE f(E) g(E) = \int_{-\sqrt{5}}^{\sqrt{5}}dE f(E) g(E)
+\int_{|E|>\sqrt{5}} dE f(E) g(E).
\label{341}
\ee
The last integral in (\ref{341}) is nonpositive, since 
$g(E) < 0$ and $f(E)=R(E) > 0$ for $|E|>\sqrt{5}$.
 Consequently, the first
integral on the right-hand side of (\ref{341}) [which is identical to the last integral on the
right-hand side of (\ref{(340)})] is nonnegative. Therefore $\tilde{I} (R) \ge \tilde{I }(P_0)$ for any function
$R(E)$ satisfying conditions (\ref{(338a)}) and (\ref{(338b)}). Thus we obtain the inequality
\begin{equation}
\tau_0\Delta E\ge\frac {3\pi}{5\sqrt {5}}\hbar\approx 
0.843\hbar.
\label{342}
\end{equation}
The equality sign in (\ref{342}) is achieved for the distribution 
$P_{*}(E) = (\Delta E)^{-1}P_0(E/\Delta E)$ (here we return to dimension variables), 
where $P_0(E)$ is given by (\ref{(339)}) and 
the energy dispersion $\Delta E$ can be arbitrary positive number (remember that the energy scale
is shifted in such a way that $\langle E \rangle=0$).
The corresponding  `nondecay probability' can be calculated with the aid of formulas
(\ref{defQ}) and (\ref{326}). It has the form
\begin{equation}
Q_*(t)= 9\left(\sin z- z\cos z\right)^2/{z^6},
\label{343}
\end{equation}
where
$ z(t)= \sqrt{5} t\Delta E/\hbar = 3\pi {t}/(5\tau_0)$.

Two other examples of energy distributions possessing products $\tau_0\Delta E$ close
to the minimal possible value (\ref{342}) were given in \cite{Gisl85}. The first of
them is the Gaussian distribution
\begin{equation}
Q(t)=\exp\left[-\pi\left(t/2\tau_0\right)^2\right]=\exp\left[-\left(t\Delta E/\hbar\right)^2\right] ,  
\label{344}
\end{equation}
\be
P(E)=\left(\Delta E \sqrt{2\pi}\right)^{-1}\exp\left[-\frac12\left(E/\Delta E\right)^2\right],
\label{344P}
\ee
\be
\tau_0\Delta E= \sqrt{\pi}/2\approx 0.886.
\label{344Et}
\ee
One can check that the nondecay probability (\ref{344}) satisfies the Luo inequality (\ref{Luo}).

The second example is the stepwise energy distribution
\begin{equation}
P(E)=\left\{\begin{array}{cc}
\sqrt{3}/(6\Delta E), &|E|\le \sqrt{3}\Delta E\\
0, &|E|> \sqrt{3}\Delta E\end{array}
\right.
\label{345}
\end{equation}
\be
Q(t)=\left[\frac {\sin\left(\sqrt{3} t\Delta E/\hbar\right)}{\sqrt{3} t\Delta E/\hbar}\right]^2, 
\label{345a}
\ee
\[
\tau_0\Delta E=\frac {\pi}{2\sqrt {3}}\approx 0.907.
\]
The time-dependences 
 (\ref{343}), (\ref{344}) and (\ref{345a}) seem to be very far from
those which can occur in real decaying physical systems. 
However, they can be interesting from the point of view of the problem of 
{\em maximal speed of quantum evolution}: see section \ref{sec-speed}.
Mathematical aspects of the time decay problem (time asymmetry) were discussed in \cite{Bohm81,Bohm03,Civi04,Bohm11,Marchetti13}.

\subsection{Modifying definitions of the decay times and energy spread}
\label{subsec-modifdecaytimes}

The energy distributions in realistic decaying systems are close to
the Lorentz distribution (\ref{328}). In these cases the decay time is determined not by the
energy dispersion (\ref{varE}), but by the energy level width $\Gamma$. 
Changing the form of the distribution function $P(E)$ at its `tail' (for $|E - E_0| \gg \Gamma$) we can change the variance
$\Delta E$ significant1y, but the decay time will remain practically unchanged. This
fact was emphasized long ago by Fock and Krylov \cite{FoKry}. 
Therefore, to fill inequality (\ref{ETh}) with a physical content in the decay problems, we must find 
a more reasonable definition of energy `uncertainty' $\Delta E$ (as was questioned by Mandelstam and Tamm), 
linking it not to the energy variance (\ref{varE}), but
to some other quantity, in such a way that it would be close to the energy level width for distributions similar to the Lorentz one. 

Several possible definitions of this kind were proposed in \cite{183}. They were based on some results of 
study \cite{Bauer}, where the concept of an
`equivalent width' of a function was introduced. Precisely, the equivalent width $W(\vf)$ of function $\vf(x)$ 
 was defined as 
\be
W(\vf) =\int_{-\infty}^{\infty} \vf(x)dx/\vf(0),
\label{(346)}
\ee
provided the integral exists and $\vf(0)\neq 0$.
Further calculations are based on the following simple observation: if two functions 
$f(x)$ and $\tilde{f}(y)$ are related by the Fourier transformation, i.e.
\[
\tilde{f}(y) =\int_{-\infty}^{\infty}e^{-ixy}f(x)dx, 
\qquad
f(x) =\int_{-\infty}^{\infty}e^{ixy}\tilde{f}(y)dy /(2\pi),
\]
then
\be
W(f) W( \tilde{f}) =2\pi.
\label{(347)}
\ee
Now let us look at the following consequence of equation (\ref{326}) 
(in dimensionless variables with $\hbar=1$):
\[ 
\int_{-\infty}^{\infty} e^{-i Et}\left[P(E)\right]^2\,
dE=\frac 1{2\pi}\int_{-\infty}^{\infty} \chi\left(t'\right)\chi^{*}\left(t'-
t\right)\,dt',
\] 
where  $P(E)$ is the energy distribution function and $\chi(t)$ is the non-decay probability amplitude.
Defining $f(E)=[P(E+E_0)]^2$, where $E_0$ is an arbitrary real number, we have   
\[
\tilde{f }(t)= \exp(iE_0 t)\int_{-\infty}^{\infty} \chi\left(t'+t\right)\chi^{*}\left(t'\right)\,dt'/(2\pi).
\]
Obviously, 
\[W(f)=\int_{-\infty}^{\infty} \left[P(E)\right]^2\, dE/[P(E_0)]^2,
\]
 whereas
\[
W(\tilde{f})= \frac{\int\int dt dt' \exp(iE_0 t) \chi\left(t'+t\right)\chi^{*}\left(t'\right)}
{\int_{-\infty}^{\infty} |\chi\left(t'\right)|^2)\,dt'}.
\]
Obviously
\[
|W( \tilde{f})| \le \frac{\int\int dt d\tau | \chi\left(\tau+t\right)\chi^{*}\left(\tau\right)|}
{\int_{-\infty}^{\infty} |\chi\left(\tau\right)|^2)\,d\tau} =
\frac{\left[\int d\tau | \chi\left(\tau\right)|\right]^2}
{\int |\chi\left(\tau\right)|^2)\,d\tau}.
\]
In addition, the identity $|W(f) W( \tilde{f})| =2\pi$ holds for any value of $E_0$ as a 
consequence of  (\ref{(347)}).
Remembering that $\chi(-t)=\chi^*(t)$ for real energy distribution function $P(E)$ and returning to
dimension variables and
the nondecay probability $Q(t) = |\chi(t)|^2$, we arrive at the inequality
\begin{equation}
\frac {\int\left[P(E)\right]^2\,dE}{\left[P(E_0)\right]^2}\cdot\frac {\left[\int_0^{\infty}\sqrt {Q(t)}\,dt\right
]^2}{\int_0^{\infty}Q(t)\,dt}\ge\pi\hbar,
\label{350}
\end{equation}
which holds for an arbitrary value $E_0$, in particular for $E_0$ corresponding to
the maximum of function $P(E)$. Looking at the left-hand side of (\ref{350}), it seems reasonable
to introduce the following definitions of the decay time and energy uncertainty:
\begin{equation}
\tau_{*}=\frac {\left[\int_0^{\infty}\sqrt {Q(t)}\,
dt\right]^2}{4\int_0^{\infty}Q(t)\,dt}, \quad
\Delta E_{*}=\frac {\int\left[P(E)\right]^2\,dE}{\max\left
[P(E)\right]^2}.
\label{351}
\end{equation}
Then $\tau_{*}=\tau$ for the exponential decay  (\ref{327}) 
(the normalization factor $1/4$ in the definition of $\tau_*$ is chosen just in order to ensure this equality).
Therefore the lower bound for the product
$\tau_* \Delta E_*$ is the same as in (\ref{332}) or (\ref{334}), but with different meanings of symbols on the left-hand
side:
\begin{equation}
\tau_{*}\Delta E_{*}\ge\pi\hbar /4.
\label{353}
\end{equation}
  On the other hand, taking into account  (\ref{333}) and (\ref{(336)}), we can
define the characteristic decay time and energy uncertainty as
\begin{equation}
\tau_{**}=\frac 12\int_0^{\infty}\sqrt {Q(t)}\,dt, \quad
\Delta E_{**}=\left[\max P(E)\right]^{-1},
\label{356}
\end{equation}
 rewriting (\ref{350}) in the form 
\begin{equation}
\tau_{**}\Delta E_{**}\ge\pi\hbar /2
\label{355}
\end{equation}
(the coefficient $1/2$ in the definition of $\tau_{**}$
is chosen again in order to ensure the equality $\tau_{**}=\tau_0$ for the exponential
decay law).
Three sets of decay times and energy uncertainties  are connected as follows:
\begin{equation}
\tau_{*}=\tau_{**}^2/\tau_0, \quad
\Delta E_{*}=\Delta E_{**}^2\tau_0/\pi\hbar.
\label{357}
\end{equation}

Note that inequality (\ref{355}) can be obtained directly from (\ref{(347)}) if one takes
$f= P(E)$. Then $\tilde{f} =\chi(t)$  according to (\ref{326}), so that (\ref{355}) follows from (\ref{(347)})
due to the inequality $W(\tilde{f}) \le W(|\tilde{f}|)$
(the value $W(\tilde{f})$ is real and positive in this case due to (\ref{(347)}), since function $f=P(E)$  is positive).

Relation (\ref{355}) becomes an equality for the exponential decay law (\ref{327}) with energy distribution (\ref{328}). 
The same is true for relation (\ref{353})
due to (\ref{357}). Therefore, inequalities (\ref{353}) and (\ref{355}) can be considered as
reasonable (and exact) energy-time uncertainty relation for decaying systems. 

Taking  $|u\rangle =|\psi(0)\rangle$) in the Wigner relations (\ref{(322)}), (\ref{(323)}) and (\ref{(324)}),
we can rewrite them in terms of functions $P(E)$ and $Q(t)$ as follows (relaxing the restriction of the
energy boundedness from below): 
\be
\tilde{\tau}^2 = \frac{\int_0^{\infty} t^2 Q(t) dt}{\int_0^{\infty}  Q(t) dt}
= \frac1{\tau_0}\int_0^{\infty} t^2 Q(t) dt,
\label{tautil}
\ee
\be
\vep^2 = \frac{\int_{-\infty}^{\infty} (E-\langle E\rangle)^2 [P(E)]^2 dE}
{\int_{-\infty}^{\infty}  [P(E)]^2 dE}, 
\quad
\langle E\rangle =\frac{\int_{-\infty}^{\infty} E [P(E)]^2 dE}{\int_{-\infty}^{\infty}  [P(E)]^2 dE},
\label{358}
\ee
\be
\vep\tilde{\tau} \ge \hbar/2.
\label{(324til)}
\ee
For the exponential decay law (\ref{327}) with the Lorentzian energy distribution (\ref{328})  we have
\[
\tilde\tau=\sqrt{2}\hbar/\Gamma, \quad \langle E\rangle=E_0, \quad \vep=\Gamma/2,
\quad \tilde\tau \vep=\hbar/\sqrt{2}.
\]

As one can see, the replacement of the weight factor $P(E)$ by $[P(E)]^2$ 
 in the formula for the `effective variance' $\vep^2$  enables
us to suppress the slowly decreasing `tail' of the Lorentz distribution function P(E) (\ref{328}). 
As a result, the `effective variance'  proves to be of the order of the
physically acceptable width of energy level.
Moreover, for the Gaussian decay law (\ref{344})-(\ref{344P}) the inequality (\ref{(324til)})
becomes strict equality (similar to the minimal product $\Delta x \Delta p=\hbar/2$ for the
Gaussian distributions in the coordinate and momentum spaces). Indeed, we obtain in this case
$\tilde\tau=\hbar/(\sqrt{2}\Delta E)$ and $\vep=\Delta E/\sqrt{2}$.

Fujiwara \cite{Fuji70} introduced one more characteristic decay time 
\be
\Delta_1t=\sqrt{\tilde\tau^2-\tau_1^2}, \quad
\tau_1= \frac{\int_0^{\infty} t Q(t) dt}{\int_0^{\infty}  Q(t) dt},
\label{(360)}
\ee
where $\tilde\tau$ is defined by equation (\ref{tautil}).
Then  we have for the exponential decay law (\ref{327})
\be
\tau_1=\Delta_1 =\hbar/\Gamma, \quad \vep\Delta_1t=\hbar/2.
\label{(361)}
\ee
However, the value $\hbar/2$ is not the lower limit for the product $\vep\Delta_1t$
 (as was mentioned without proof and details in \cite{Fuji70}). It is clear from the example of Gaussian states
 (\ref{344})-(\ref{344P}): 
\[
\tau_1 = \hbar/(\Delta E \sqrt{\pi}), \quad
\Delta_1 t=(\hbar/\Delta E)\sqrt{(\pi-2)/(2\pi)}, 
\]
\be
(\vep\Delta_1t )_{Gauss}= \hbar\left[(\pi-2)/(4\pi)\right]^{1/2} \approx 0.301\hbar.
\label{292Et}
\ee
Other definitions of the `evolution time' were considered in \cite{Boykin07}.
Concrete calculations of functions like $\xi_u(t)$ and $\eta(E)$ (\ref{(318)}), as well as the
related
quantities $\tilde\tau$, $\vep$ and $\tau_1$, for different choices of the
reference states $|u\rangle$ and solutions to the Schr\"odinger equation $\psi(t)$ 
for a partiele in a uniform electric field (when the spectrum of the Hamiltonian is
continuous and extends from $-\infty$ to $+\infty$) were made in study \cite{Fuji80}.

Hilgevoord and Uffink \cite{Hilg90,Uffink93} introduced a measure for the uncertainty in energy
$W_{\alpha}(E)$ as the size of the shortest interval $W$ such that
\be
\int_{W} |\langle E|\psi(0)\rangle|^2 dE = \alpha.
\ee
The time extension $\tau_{\beta}$ was defined as the minimal time it takes for $|\psi (0)\rangle$ to evolve to a state 
$|\psi(\tau)\rangle$ such that
\be
|\langle \psi(0)|\psi(\tau)\rangle| =\beta.
\ee
It appears that 
\be
\tau_{\beta}W_{\alpha} \ge 2\hbar\, \mbox{arccos}\left(\frac{\beta+1-\alpha}{\alpha} \right) \;\; \mbox{for}
\;\; \beta \le 2\alpha -1.
\ee
In particular, one obtains for the Lorentzian distribution (\ref{328}) the values
\[
W_{\alpha}=\Gamma \tan(\alpha\pi/2), \quad \tau_{\beta} =4\tau\ln(1/\beta).
\]

\section{`Time--frequency' uncertainty relations}
\label{sec-freq}

It appears that the problem of finding the lower limit for the product $\vep\Delta_1t $ is closely related to the {\em time--frequency\/}
uncertainty relation. Therefore it is worth discussing this problem in details. 

Dividing both sides of inequality (\ref{ETh}) by the Planck constant $\hbar$ and using the
Planck formula $E=\hbar\omega$ one arrives at the approximate inequality
\be
\Delta\omega \Delta t  \gtrsim 1
\label{271}
\ee
where the right-hand side can be replaced with the same accuracy by $\pi$ or $2\pi$.
Actually, this inequality, which
connects some effective duration of signal $\Delta t$ with some effective spectral width $\Delta\omega$,
was known in optics and radiotechnics long before the birth of quantum mechanics.
Moreover, it was frequently used for illustrations of the quantum-mechanical uncertainty relations
at the early years of quantum mechanics.
However, exact formulations of relation (\ref{271}) were given after strict formulations of the
quantum-mechanical UR only. By analogy with the strict Heisenberg--Weyl UR $\Delta x \Delta p \ge \hbar/2$,
some people believe that the precise version of (\ref{271}) is the inequality
\be
\Delta\omega \Delta t  \ge 1/2,
\label{272}
\ee
where the quantities $\Delta\omega$ and $ \Delta t$ are defined in the same standard way as in quantum mechanics:
\be
(\Delta t)^2= \int_{-\infty}^{\infty}(t-\overline{t})^2 f^2(t) dt, \quad
\overline{t}=\int_{-\infty}^{\infty}t f^2(t) dt, 
\label{273t}
\ee
\be
(\Delta \omega)^2= \int_{-\infty}^{\infty}(\omega-\overline{\omega})^2 |F(\omega)|^2 d\omega, 
\ee
\be 
\overline{\omega}=\int_{-\infty}^{\infty}\omega |F(\omega)|^2 d\omega,  
\label{273om}
\ee
\be
f(t)=\frac1{\sqrt{2\pi}}\int_{-\infty}^{\infty}F(\omega) e^{i\omega t} d\omega, 
\label{fFt}
\ee
\be
F(\omega)= \frac1{\sqrt{2\pi}}\int_{-\infty}^{\infty} f(t) e^{-i\omega t} dt.
\label{fF}
\ee
It is supposed here that the signal $f(t)$ is normalized according to relations 
\be
\int_{-\infty}^{\infty}f^2(t)dt= \int_{-\infty}^{\infty}|F(\omega)|^2)d\omega=1.
\label{norm}
\ee
In such a case, although inequality (\ref{272}) is correct, it is useless in many practical
situations. The problem is that, in contrast to the quantum-mechanical case,
the time-dependent signal $f(t)$ is always a {\em real function}. Therefore
\be
F(\omega)=F^*(-\omega), \quad |F(\omega)|^2= |F(-\omega)|^2,
\label{275}
\ee
so that $\overline\omega \equiv 0$ for any signal. This result contradicts our intuition.
For example, for any narrow-band signal, function $|F(\omega)|^2$ has two
narrow peaks, centered symmetrically with respect to point $\omega=0$.
But in such a case the quantity $\Delta\omega$, defined by equation (\ref{273om}),
characterizes not the spectral width of the signal, but its carrying frequency.

Therefore, in order to have a meaningful UR for real signals, it seems much more natural to 
take into account only positive frequencies and redefine
the average frequency and spectral width as follows:
\be
(\Delta \omega_+)^2= 2\int_{0}^{\infty}(\omega-\overline{\omega}_+)^2 |F(\omega)|^2 d\omega, 
\label{276}
\ee
\be
\overline{\omega}_+ =2\int_{0}^{\infty}\omega |F(\omega)|^2 d\omega,  
\label{276+}
\ee
where the factor $2$ appears due to the normalization.
Thus we arrive at the problem of finding the minimal possible value of the product $\Delta t \Delta \omega_+$.
This problem was analyzed long ago (in 1934) by Mayer and Leontovich \cite{Mayer34}. However, this study was practically
unknown for a long time. More known was the paper by Gabor \cite{Gabor46}, where different definitions of the
frequency and duration of signals were considered. Gabor introduced the concept of `analytical signal': a complex
function $f_{+}(t)$ whose Fourier transform does not contain negative frequencies,
\be
f_{+}(t)= \int_0^{\infty}F(\omega) e^{i\omega t} d\omega/\sqrt{\pi},
\label{277-1}
\ee
\be
F_{+}(\omega) =\int_{-\infty}^{\infty}f_{+}(t) e^{-i\omega t} {dt}/{\sqrt{2\pi}}
=
\left\{
\begin{array}{cc}
\sqrt{2}F(\omega), & \omega \ge 0
\\
0, & \omega <0
\end{array}
\right.
\label{277-2}
\ee
Putting the function $F_{+}(\omega)$ in the definitions (\ref{273om}), we have the equalities
$\overline\omega=\overline\omega_{+}$ and $\Delta\omega=\Delta\omega_{+}$. Therefore Gabor suggested
to replace inequality (\ref{272}) by the following modification:
\be
\Delta\omega_{+} \Delta t_{+}  \ge 1/2,
\label{278}
\ee
where
\[
(\Delta t_{+})^2= \int_{-\infty}^{\infty}(t-\overline{t}_{+})^2 |f_{+}(t)|^2 dt, 
\qquad
\overline{t}_{+}=\int_{-\infty}^{\infty}t |f_{+}(t)|^2 dt.
\]
Of course, inequality (\ref{278}) is correct. But now we meet another problem: what is the physical meaning
of the quantities $\Delta t_{+}$ and $\overline{t}_{+}$, and how these new quantities are related to $\Delta t$ 
and $\overline{t}$?
 It was shown by E. Wolf \cite{Wolf58} that $\overline{t}_{+}=\overline{t}$ and
$\Delta t_{+}=\Delta t$ under the condition
\be
F(0)= \frac1{\sqrt{2\pi}}\int_{-\infty}^{\infty} f(t)  dt  =0.
\label{280}
\ee
In such a case
\be
\Delta\omega_{+} \Delta t  \ge 1/2,
\label{281}
\ee
The proof of (\ref{281}) was given by Kay and Silverman \cite{KaySilv57} without using the concept of analytical signal.
We reproduce it (in a slightly modified form) in the case of $\overline{t}=0$, which can be always achieved by a shift of the
origin of the time axis.

Remembering that the Fourier transform of function $tf(t)$ is $idF/d\omega=iF'(\omega)$, we can write
(using the Parseval identity)
\be
(\Delta t)^2(\Delta\omega_{+})^2 =4\int_0^{\infty}|F'(\omega)|^2 d\omega 
\int_0^{\infty}
\left(\omega - \overline{\omega}_+\right)^2 |F(\omega)|^2 d\omega.
\label{282}
\ee
Using the Schwartz inequality (\ref{(283)}) with $f(\omega)=F'(\omega)$ and 
$g(\omega)= \left(\omega - \overline{\omega}_+\right) F(\omega)$ we are led to the
following chain of relations:
\beqn
\fl
\Delta\omega_{+} \Delta t   &\ge &
\left|\int_0^{\infty} \left(\omega - \overline{\omega}_+\right) \left[F'F^* +(F')^* F\right] d\omega\right|
=
\left| \int_0^{\infty} \left(\omega - \overline{\omega}_+\right)\frac{d}{d\omega}|F(\omega)|^2 d\omega\right|
\nonumber \\  \fl &=&
\left| \left.\left\{\left(\omega - \overline{\omega}_+\right) |F(\omega)|^2\right\}\right\vert_0^{\infty}
- \int_0^{\infty}|F(\omega)|^2 d\omega \right|
=
\left| 1/2 -|F(0)|^2 \overline{\omega}_+\right|.
\label{284}
\eeqn
However, not only the equality in (\ref{284}) cannot be achieved (as was shown in \cite{KaySilv57}), but it gives
in many cases the lower bound on the product $\Delta\omega_{+}\Delta t$, which is significantly smaller than really
admissible values.

For example, let us consider the Gaussian signal with $\overline{t}=0$
[that does not satisfy condition (\ref{280})]:
\be
f(t)=\left(\sigma^2/\pi\right)^{1/4} \exp\left(-\sigma^2 t^2/2\right),
\label{fG}
\ee
\be
F(\omega)= \left(\pi\sigma^2\right)^{-1/4} \exp\left(-\omega^2/2\sigma^2 \right).
\label{FG}
\ee
Then
\[
\Delta t= \frac1{\sigma\sqrt{2}}, \quad
\overline{\omega}_{+}= \frac{\sigma}{\sqrt{\pi}}, \quad
\Delta\omega_{+}= \sigma\left[\frac{\pi-2}{2\pi}\right]^{1/2},
\]
so that {\em for any\/} $\sigma$ we obtain the same value of the product
\be
(\Delta t \Delta\omega_{+})_{Gauss}= \left[(\pi-2)/(4\pi)\right]^{1/2} \approx 0.301,
\label{292}
\ee
whereas $\frac12\left| 1-2|F(0)|^2 \overline{\omega}_+\right|=\frac12\left| 1-2/\pi\right| \approx 0.18$.

The exact lower bound for the product $\Delta t \Delta\omega_{+}$ was found by Mayer and Leontovich \cite{Mayer34}, whose
results were reproduced in the book \cite{111}. However, these studies were unknown in the West for a long time. 
Hilberg and Rothe \cite{Hilberg71} gave in 1971 a solution to the problem, very similar to the Mayer--Leontovich approach (except for the final stage).
Then Borchi and Pelosi \cite{Borchi80} gave a derivation, using a quantum-mechanical analogy with a double harmonic oscillator.
We reproduce below the Mayer--Leontovich approach, following \cite{111}. The reason is that this nice solution is still not well known.
Besides, we give some details absent in \cite{Mayer34,111} and compare their results with \cite{Hilberg71,Borchi80}.

We assume that $\overline{t}=0$, since this equality can be achieved by a simple shift in time.
It was proven in \cite{Mayer34} that
the minimum of product $\Delta\omega_{+} \Delta t $ can be achieved under the condition $\overline{t}=0$ for {\em even\/}
functions of time $f(t)$. Then the spectral function $F(\omega)$ is also even and real 
(so that condition (\ref{280}) is not satisfied).
Abandoning for a while the normalization condition on function $F(\omega)$, one
can rewrite the right-hand side of (\ref{282}) as
a functional $\Omega\{F\}$ of the function $F(\omega)$:
\be
\Omega\{F\} =AN_2/N_0^2 - AN_1^2/N_0^3,
\label{286}
\ee
where
\[
A = 2\int_0^{\infty}[F'(\omega)]^2 d\omega, \quad
N_m =2\int_0^{\infty} \omega^m F^2(\omega) d\omega.
\]
The variation $\delta\Omega$ must go to zero for the extremal function. This means that
\beqn
N_0^4 \delta\Omega &=& N_0 \left(N_0 N_2 -N_1^2\right)\delta A 
+N_0 A \left(N_0 \delta N_2 - 2N_1 \delta N_1\right) 
\nonumber \\ &&
+A \left( 3N_1^2 -2N_0 N_2\right)\delta N_0 =0.
\label{288}
\eeqn
Let us introduce the notation $a=A$, $b=N_2$ and $c=N_1$ {\em for the extremal function}.
Imposing now the normalization condition $N_0=1$
(then $c \equiv \overline\omega_{+}$), we can represent equation (\ref{288}) as
\be
\delta \int_0^{\infty}\Big\{  a\left(\omega^2 -2c\omega +3c^2 -2b\right) F^2(\omega) 
+ \left(b-c^2\right)[F'(\omega)]^2 \Big\} d\omega =0.
\label{289}
\ee
This is a standard variation problem. The corresponding Euler--Lagrange equation reads
\be
 \left(b-c^2\right) F''(\omega) - a\left(\omega^2 -2c\omega +3c^2 -2b\right) F =0.
\label{290}
\ee
Multiplying equation (\ref{290}) by $F'(\omega)$ and integrating from $0$ to $\infty$ we get
\[
\left. a\left(\omega^2 -2c\omega +3c^2 -2b\right) F^2(\omega)\right|_0^{\infty}
= \left(b-c^2\right)\left.[F'(\omega)]^2\right|_0^{\infty} =0,
\]
because $F'(0)=0$ due to the parity of function $F(\omega)$.
Consequently $b=3c^2/2$. 
Introducing the notation
\[
\Omega_{min} \equiv \left[(\Delta t \Delta\omega_{+})_{min}\right]^2 \equiv \mu^2
\]
we obtain
\be
 \mu^2 = \frac13 ab =
\frac43 \int_0^{\infty}[F'(\omega)]^2 d\omega
\int_0^{\infty} \omega^2 F^2(\omega) d\omega.
\label{Om-mu}
\ee
Then the Schwartz inequality yields the lower bound
\be
\mu  \ge \frac{2}{\sqrt{3}}
\left|\int_0^{\infty} \omega F'(\omega) F(\omega) d\omega \right| =\frac1{\sqrt{12}}
\approx 0.289.
\label{291}
\ee
In view of (\ref{292}) we can conclude that $0.289 \le \mu \le 0.301$
(the result $\mu \approx 0.3$ was suggested also in \cite{KaySilv57}).
To find the exact value of $\mu$, it is necessary to solve equation (\ref{290}),
which can be reduced to the known Weber equation \cite{Bateman}
\be
d^2 F/dz^2 +\left(\nu +1/2 -z^2/4\right)F =0
\label{293}
\ee
with (taking into account the relation $b=3c^2/2$)
\[
\nu +1/2 =\sqrt{a b/3} \equiv \mu, \quad
z(\omega)=2\sqrt{\mu}(\omega/c -1).
\]

Solutions to equation (\ref{293}) that are limited for $z\to\infty$ are the parabolic cylinder functions
$D_{\nu}(z)$. Consequently, the extremal frequency spectrum is given by the formula
\be
F_{\mu}(\omega) =D_{\mu-1/2}\left(2\sqrt{\mu}[\omega/c -1]\right).
\label{FD}
\ee 
(Hilberg and Rothe \cite{Hilberg71} did not want to analyze this solution, choosing a longer way.)
The extremal function $F_{\mu}(\omega)$ must satisfy the additional restriction
$F_{\mu}'(0)=0$. Since $z(0)= -2\sqrt{\mu}$, the value of $\mu$
can be found as the solution of equation
\be
D'_{\mu-1/2}\left(-2\sqrt{\mu}\right)=0, \quad D'_{\nu}(z) \equiv dD_{\nu}(z)/dz.
\label{294}
\ee
Using the integral representation of the parabolic cylinder functions with negative index $\nu<0$ \cite{Bateman}
(in our case $\mu \approx 0.3$, so that $\nu \approx -0.2$)
\[
D_{\nu}(z)=\frac{1}{\Gamma(-\nu)}
\int_0^{\infty} \exp\left(-z^2/4
-zt -t^2/2\right) t^{-\nu-1} dt
\]
one can reduce equation (\ref{294}) to \cite{Mayer34}
\be
\int_0^{\infty} \exp\left(2\rho t -t^2/2\right) t^{-\rho^2-1/2} (t-\rho) dt =0,
\label{295}
\ee
where $\rho \equiv \sqrt{\mu}$.
However, Mayer and Leontovich did not give a numerical solution to this equation, having stopped at this point.
Our numerical solution of equation (\ref{295}) gives the value
\be
\mu \equiv \left(\Delta\omega_{+} \Delta t\right)_{min}= 0.29505306...
\label{296}
\ee
The value given by 
Hilberg and Rothe \cite{Hilberg71} was (one should divide their result $1.180 212...$ by the factor $4$ due to
the different definition of `uncertainties')  $0.295053$, and the same value was given by 
Borchi and Pelosi \cite{Borchi80}, who solved equation (\ref{294}) using another way.

\begin{figure}[htb]
\par
\vspace{-8mm}
\begin{center} 
\includegraphics[width=130mm]{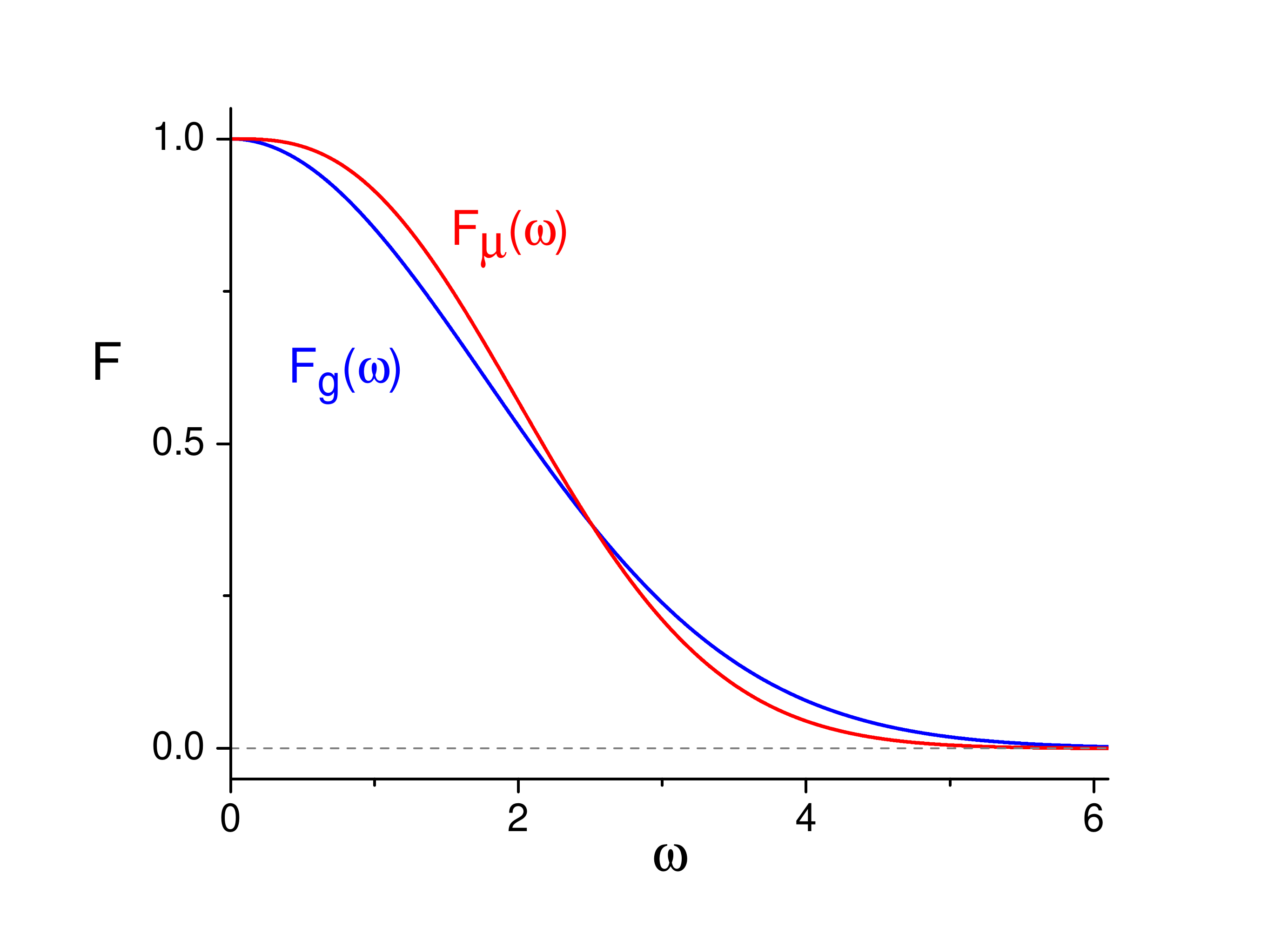}
\end{center} 
\par
\vspace{-8mm}
\caption{
Gaussian spectral function $F_g(\omega)$  (\ref{FG}) and the `minimal uncertainty product' spectral function $F_{\mu}(\omega)$  (\ref{FD})
with equal dimensionless average values $\overline\omega_{+}=1$. 
}
\label{fig-Fom}
\end{figure}

\begin{figure}[htb]
\par
\vspace{-8mm}
\begin{center} 
\includegraphics[width=130mm]{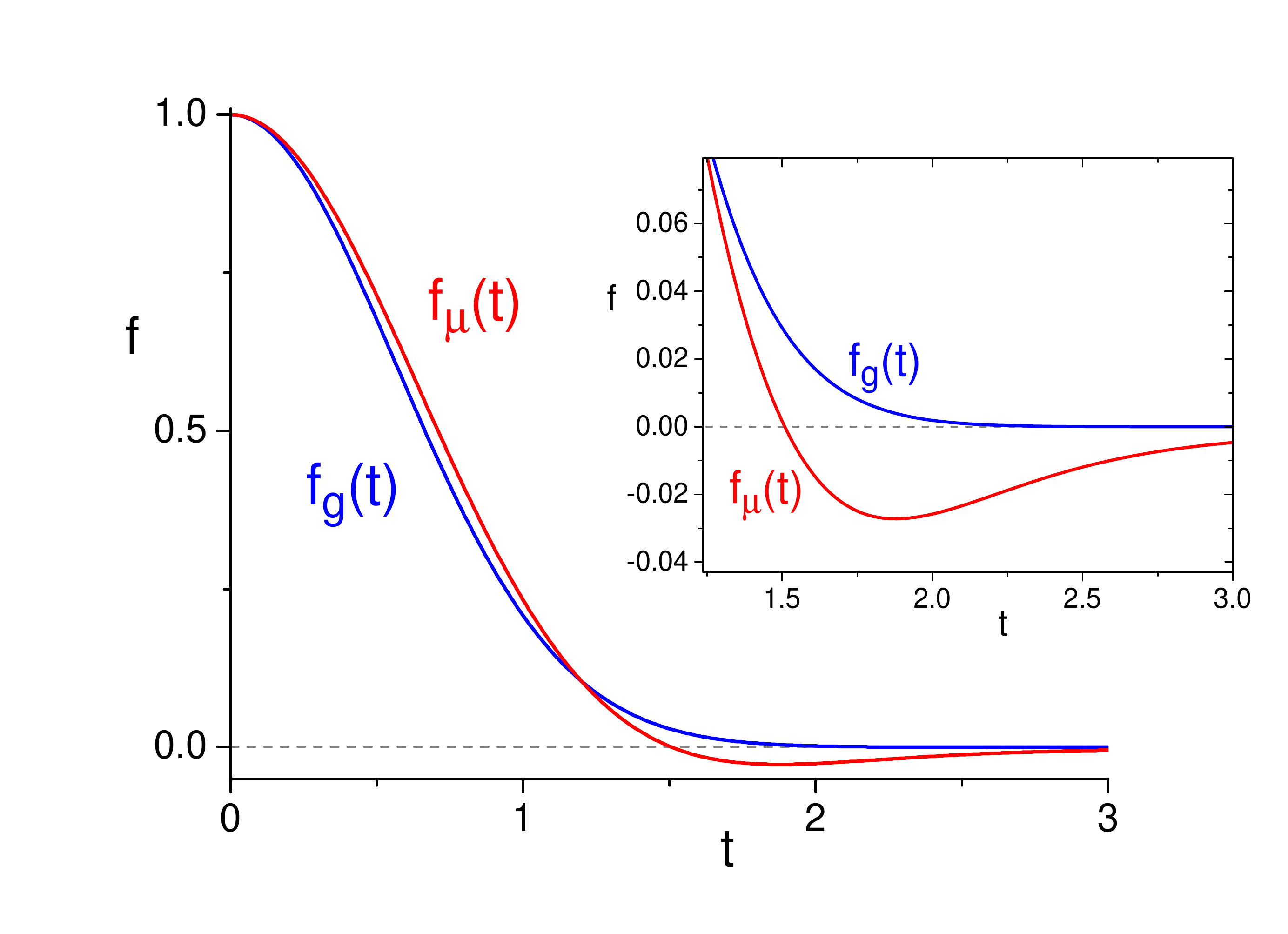}
\end{center} 
\par
\vspace{-8mm}
\caption{Time-dependent signals corresponding to spectral functions  (\ref{FG}) and   (\ref{FD})
versus dimensionless time variable $\overline\omega_{+} t$ with $\overline\omega_{+}=1$.
}
\label{fig-ft}
\end{figure}

We see that Gaussian signals (\ref{fG})-(\ref{FG})  possess the value of product
$\Delta\omega_{+} \Delta t$ that is very close to the minimal possible value (\ref{296})
(this was noticed in \cite{111}). Therefore it is interesting to compare Gaussian packets with the 
`minimal time-frequency uncertainty packets' (\ref{FD}). We notice that the
extremal frequency spectrum is more `flat' near the point $\omega=0$, since the consequence
of equation (\ref{290}) with $b=3c^2/2$ is the equality $F''(0)=0$. On the other hand,
the extremal function decays slightly faster than the Gaussian signal for
$\omega\to\infty$, since asymptotically $D_{\nu}(z) \sim z^{\nu}\exp\left(-z^2/4\right)$ \cite{Bateman}.
The comparison of the spectral functions  (\ref{FG}) and  (\ref{FD}) with identical values of $ \overline\omega_{+}$
 is shown in figure \ref{fig-Fom}.
Both the functions are normalized by their values at $\omega=0$, and the 
dimensionless variable $\omega$ means in fact the ratio $\omega/ \overline\omega_{+}$. Formally, this means that
we take $c=1$ in formula (\ref{FD}) and $\sigma=\sqrt{\pi}$ in (\ref{FG}).
Figure \ref{fig-ft} shows the time-dependent real Gaussian signal (\ref{fG}) and the real Fourier transform
$f_{\mu}(t)$
of spectral function  (\ref{FD}) as functions of the dimensionless variable  $\overline\omega_{+} t$.
[For $\omega<0$ we use the relation $F(-\omega)=F(\omega)$.]
Again, both functions are normalized by their values at $t=0$. 
We see that $f_{\mu}(t)$ can assume negative (although relatively small) values.
Similar figures were given in \cite{Hilberg71}, but the `extra-flatness' of function $F_{\mu}(\omega)$ at its top
was not noticed there.

 Approximate `minimal time-frequency uncertainty packets' in the form of truncated Gaussian packets
$F(\omega)=\exp\left[-(|\omega| -m)^2/2\sigma^2\right]$ were studied in \cite{KaySilv57},
where it was shown that $\Delta\omega_{+} \Delta t \approx 0.3$ for certain choices
of parameters $m$ and $\sigma$. 
Spectral functions in the form of a combination of two Gaussian functions centered at the points $\pm\omega_1$
were considered in \cite{Borchi81}. The minimal value of the product $\Delta t \Delta\omega_{+}$ for such functions
is $0.29535$, but the functions $f(t)$ have the form of Gaussian packets multiplied by $\cos(\omega_1 t)$, so these
functions also assume small negative values in some time intervals.
Parabolic cylinder functions arise in the problem of minimization of the product $\Delta t \Delta\omega$ in the case
of band-limited spectral functions satisfying the condition $F(\omega)\equiv 0$ for $|\omega|>\Omega$ \cite{Borchi85}.
Some results of recent studies on the time--frequency problem
can be found in \cite{Flandrin98,Shinde01,Cohen04,Xu10,Dang13}.

\subsection{Consequences for the $\Delta E \Delta t$ problem}

Now we can return to the $\Delta E \Delta t$ problem. Comparing formulas (\ref{273t})-(\ref{fF}) with
(\ref{defQ}), (\ref{326}), (\ref{tautil}) and (\ref{358}), one can see that the quantities $\vep$ (\ref{358}) and
$\Delta_1t$ (\ref{(360)}) are equivalent to $\Delta t$ (\ref{273t}) and $\Delta\omega_{+}$ (\ref{276}), 
respectively.\footnote{It is funny that the time variable in the frequency--time problem should be substituted by the energy 
in the time--energy problem, while the frequency should be substituted by time.}
Therefore, the lower bound of the product $\vep \Delta_1t$ cannot be less than  $0.29505\hbar$. This boundary
could be attained by replacing function $[f(t)]^2$ with $[P(E)]^2$ and $[F(\omega)]^2$ with 
\[
Q_*(t)=[\chi_*(t)]^2= G \left\{D_{\mu-1/2}\left(2\sqrt{\mu}[t/c -1]\right)\right\}^2,
\]
where the constant factor $G$ ensures the fulfillment of the condition $Q(0)=1$.
But here we meet a difficulty: the energy distribution function $P(E)$ must be non-negative, whereas the extremal
function $f(t)$ in the frequency--time problem can assume negative values, according to figure \ref{fig-ft}.
Consequently, function $Q_*(t)$ cannot be a true non-decay probability. Therefore the minimal possible value
of the producr $\vep \Delta_1t $ is between the values $0.295\hbar$ and $0,301\hbar$ (attained for the Gaussian non-decay
probability function). The difference is rather small, but the question on the exact lower bound is still open, although, perhaps, it is not 
very important for applications.

\section{Mixed quantum states: the Eberly--Singh approach and its generalizations}
\label{sec-Eb}

An interesting generalization of the Mandelstam--Tamm inequality (\ref{(299)}) 
was given by Eberly and Singh \cite{EberSingh}. 
 It was further developed in \cite{blackbook,Kurm-ETprep}.
The generalization given there is based on the description of the evolution of a closed quantum system in the framework
of the quantum Liouville equation for the statistical operator (density matrix) $\hat\rho$
\be
i\hbar \partial\hat\rho/\partial t = [\hat{H},\hat\rho].
\label{(300)}
\ee
Remember that the notion of statistical operator was introduced by Landau \cite{Land27} and a few months later
in a more elaborated form by von Neumann \cite{Neum27}.
It is worth noting that equation (\ref{(300)}) is written in the  Schr\"odinger picture. In the Heisenberg picture it reads
$d\hat\rho/dt=0$, meaning that the statistical operator is the integral of motion (this fact was emphasized in \cite{Blanchard}).

Following \cite{blackbook} let us introduce ``the stationarity time'' $T_0$ according to the formula
\be
T_0^{-2} = \mbox{Tr}\left[\left(\partial\hat\rho/\partial t\right)^2\right].
\label{(301)}
\ee
Taking $\hat{A} = \hat{H}$ and $\hat{B}= \partial\hat\rho/\partial t \equiv \hat{\dot\rho}$ in inequality (\ref{(3)}),
we have (assuming $\hbar=1$ in order to simplify formulas; the place of $\hbar$ can be easily recovered
from the dimensionality analysis)
\be
(\Delta H)^2 \mbox{Tr}\left(\hat\rho \hat{\dot\rho}^2\right) \ge \frac14
\left|\mbox{Tr}\left(\hat\rho [\hat{H}, \hat{\dot\rho}]\right)\right|^2
= \frac14\left|\mbox{Tr}\left(\hat\rho [\hat{H},[\hat{H}, \hat{\rho}]]\right)\right|^2.
\label{(302)}
\ee
Using the identity $\mbox{Tr}\left( [\hat{A},\hat{B}]\hat{C}\right) \equiv 
\mbox{Tr}\left( [\hat{C},\hat{A}]\hat{B}\right)$,
we can rewrite (\ref{(302)}) as 
\be
(\Delta H)^2 \mbox{Tr}\left(\hat\rho \hat{\dot\rho}^2\right) \ge \frac{1}{4}
\left[\mbox{Tr}\left( \hat{\dot\rho}^2\right)\right]^2.
\label{(304)}
\ee
One can easily check the following identities:
\beqn
\mbox{Tr}\left(\hat\rho \hat{\dot\rho}^2\right) &=& \mbox{Tr}(\hat\rho [\hat\rho,\hat{H}][\hat{H}, \hat\rho])
= \mbox{Tr}(\hat\rho^3 \hat{H}^2 -\hat\rho^2 \hat{H} \hat\rho \hat{H})
\nonumber \\ 
&=& \frac12 \mbox{Tr}\left(\hat\rho^2 [\hat{H},[\hat{H}, \hat\rho]]\right)
\equiv \frac12 \mbox{Tr}\left(\hat\rho^2 \hat{A}\right).
\label{(305)}
\eeqn
Let $\{|n\rangle\}$ be a basis where
 operator $\hat\rho$ is diagonal:
\[
\hat\rho =\sum_n p_n |n\rangle\langle n|, \quad 0\le p_n \le 1, \quad \sum p_n =1.
\]
Then 
$
 \mbox{Tr}(\hat\rho^2 \hat{A})= \sum_n p_n^2 \langle n|\hat{A}|n\rangle$.
Since the quantity $\mbox{Tr}(\hat\rho \hat{\dot\rho}^2)$
is nonnegative for any statistical operator $\hat\rho$,  all mean values 
$\langle n|\hat{A}|n\rangle$  are nonnegative, as well. Then
\be
 \mbox{Tr}(\hat\rho^2 \hat{A}) \le \sum_n p_n \langle n|\hat{A}|n\rangle
=  \mbox{Tr}(\hat\rho \hat{A}).
\label{(308)}
\ee
Consequently, relations (\ref{(305)})-(\ref{(308)})
lead to the inequality
\be
\mbox{Tr}\left(\hat\rho \hat{\dot\rho}^2\right) \le \frac12 
\mbox{Tr}\left(\hat\rho [\hat{H},[\hat{H}, \hat\rho]]\right)
= \frac12 \mbox{Tr}\left(\hat{\dot\rho}^2\right).
\label{(309)}
\ee
Putting it into the left-hand side of inequality (\ref{(304)}), we arrive at the relation
(recovering the Planck constant)
\be
(\Delta H)^2 \ge \frac{\hbar^2}{2}\mbox{Tr}\left(\hat{\dot\rho}^2\right).
\label{(310)}
\ee
Defining the energy ``uncertainty'' as $\Delta E = \Delta H$,  we obtain the inequality
\be
\Delta E T_0 \ge \hbar/\sqrt{2}.
\label{(311)}
\ee

We can introduce a set of ``stationarity times'' $T_n$ according to the formula
\be
T_n^{-2} = \mbox{Tr}\left[\hat\rho^n \left(\partial\hat\rho/\partial t\right)^2\right].
\label{(312)}
\ee
They satisfy inequalities
\be
T_n \ge T_{n-1} \ge \cdots \ge T_1 \ge \sqrt{2} T_0,
\label{(313)}
\ee
following from (\ref{(308)}) and (\ref{(309)}).
In the case of pure quantum states, when $\hat\rho^n =\hat\rho$, all these inequalities turn
into strict equalities. Relations (\ref{(313)}) and (\ref{(311)}) result in the inequalities
\be
\Delta E T_n \ge \hbar, \quad n \ge 1
\label{(314)}
\ee
(namely the special case of $n = 1$ was considered for the first time by Eberly and Singh \cite{EberSingh}).

It can be interesting to see explicitly, why the times $T_0$ and $T_1$ differ by the factor $\sqrt{2}$ in the 
case of pure quantum states $\hat\rho=|\psi\rangle\langle\psi|$. The Schr\"odinger equation (with $\hbar=1$)
$|\dot\psi\rangle = -i\hat{H}|\psi\rangle$ leads to the following operator $\hat{\dot\rho}^2$:
\be
\fl
\hat{\dot\rho}^2 = |\psi\rangle\langle\psi|\hat{H}^2|\psi\rangle\langle\psi|
-\hat{H}|\psi\rangle\langle\psi|\hat{H}|\psi\rangle\langle\psi|
+
\hat{H}|\psi\rangle\langle\psi|\hat{H} - |\psi\rangle\langle\psi|\hat{H}|\psi\rangle\langle\psi|\hat{H},
\label{rhodot2}
\ee
so that $\mbox{Tr}\left(\hat{\dot\rho}^2\right) =2 (\Delta E)^2$. On the other hand, multiplying both sides
of identity (\ref{rhodot2}) by the operator $|\psi\rangle\langle\psi|$ from the left, one can see that
the last two terms in  (\ref{rhodot2}) cancel each other. Then 
\[
\hat\rho \hat{\dot\rho}^2 = |\psi\rangle\langle\psi|\hat{H}^2|\psi\rangle\langle\psi|
- |\psi\rangle\left(\langle\psi|\hat{H}|\psi\rangle\right)^2\langle\psi|
\]
and $\mbox{Tr}\left(\hat\rho\hat{\dot\rho}^2\right) = (\Delta E)^2$.

The advantage of inequalities (\ref{(311)}) and (\ref{(314)}) over the Mandelstam-Tamm
inequality (\ref{(299)}) is that the ``stationarity times'' $T_n$ are determined completely 
by the density matrix of the system alone, whereas the `duration' $\Delta t_A$ in (\ref{(299)}) 
depends on the choice of an auxiliary operator $\hat{A}$.
A generalization of the Eberly--Singh inequality in the form
\be
T_{S}^{LK} \Delta H \ge \hbar/2, \quad T_{S}^{LK}=\frac{\sqrt{\langle \hat{s}^2\rangle}}{|\langle d\hat{s}/dt\rangle|}, 
\quad \hat{s} =d\hat\rho/dt
\ee
was obtained in \cite{Leubner85};  it was shown that $T_{S}^{LK}\le T_1$, with an equality attained for pure quantum states.

\subsection{Influence of quantum purity}

Note that relations (\ref{(311)}) and (\ref{(314)}) become identities for any pure quantum state.
Therefore it is interesting to know, in particular, how the lower bound of the product $\Delta E T_0$ can depend 
on the quantum purity $\mu=\mbox{Tr}(\hat{\rho}^2)$. This problem was not considered earlier.
A simple solution can be found in the special case of a harmonic oscillator and the initial {\em Gaussian\/} states
(which remain Gaussian during the evolution). In this case it is convenient to use the description of quantum
states in terms of the real Wigner function \cite{Wig32}, related to the complex matrix elements $\langle x|\hat\rho|y\rangle$
of statistical operator $\hat\rho$ in the coordinate representation as follows,
\begin{equation}
W(q,p)=\int dv e^{ipv/\hbar}\left\langle q-{v}/{2}\right\vert \hat{\rho}\left\vert
q+{v}/{2}\right\rangle , 
\label{defWig}
\ee
\be
\int W(q,p) dq dp/(2\pi \hbar) =1.
\end{equation}
The advantage of using the Wigner function for (closed) quantum systems with quadratic Hamiltonians is that
the evolution equation coincides with the classical Liouville equation, containing partial derivatives of the first order only. 
In particular, equation (\ref{(300)}) with Hamiltonian $\hat{H}= \hat{p}^2/(2m) +m\omega^2\hat{x}^2/2$
  can be transformed to the form
\be
\frac{\partial W}{\partial t}= m\omega^2 q \frac{\partial W}{\partial p} - \frac{p}{m} \frac{\partial W}{\partial q}.
\ee
Using the formula \cite{Takabaysi54,LeeScully80,DMR80,OConn81}
\be
\mbox{Tr}\left(\hat{A}\hat{B}\right)= \int W_A(q,p)W_B(q,p) dq dp/(2\pi\hbar),
\ee
where $W_{A,B}(q,p)$ are the Wigner--Weyl symbols of operators $\hat{A}$ and $\hat{B}$,
defined in the same way as in equation (\ref{defWig}), we can rewrite definition (\ref{(301)})
as 
\be
T_0^{-2} = \int \left({\partial W}/{\partial t}\right)^2 dq dp/(2\pi\hbar).
\ee
Let us consider the initial quantum state described by the most general Gaussian Wigner function
\cite{DKM80}
\be
\fl
W(q,p) =\frac{\hbar}{\sqrt{D}} \exp\left\{ -\frac1{2D}\left[
\sigma_p\left(q-\overline{q}\right)^2 +\sigma_q \left(p-\overline{p}\right)^2 
-2\sigma_{qp}\left(q-\overline{q}\right)\left(p-\overline{p}\right)
\right] \right\}, 
\label{WG}
\ee
where
\be
D \equiv \sigma_q \sigma_p -\sigma_{qp}^2 = [(\hbar/(2\mu)]^2 \ge \hbar^2/4.
\label{d-mu}
\ee
The last inequality in (\ref{d-mu}) is the Robertson--Schr\"odinger uncertainty relation, which must
hold for any pure \cite{Rob-un,Schr-un} or mixed \cite{DKM80} quantum state.

Calculating Gaussian integrals one can obtain the following formulas:
\be
T_0^{-2}= \frac{\hbar\left(B-2D\omega^2\right)}{8D^{3/2}}, \quad 
(\Delta E)^2= \frac{B}{2} -\frac14 (\hbar\omega)^2,
\ee
\be
2\left(\Delta E T_0/\hbar\right)^2 = \left(\frac{4D}{\hbar^2}\right)^{3/2}
\left(1 + \omega^2\frac{2D-\hbar^2/2}{B-2D\omega^2}\right),
\label{2E2T2}
\ee
\[
B =  m^2\omega^4\left(\sigma_q^2 +2\sigma_q \overline{q}^2\right) +
\left(\sigma_p^2  +2\sigma_p \overline{p}^2\right)/m^2 
+2\omega^2 \left(\sigma_{qp}^2 +2\sigma_{qp} \overline{q}\overline{p}\right) .
\]
For pure Gaussian states  the product (\ref{2E2T2}) equals unity for any choice
of parameters (satisfying the restriction $D=\hbar^2/4$ under which the Wigner function
(\ref{WG}) has sense \cite{DKM80,Man08}). This shows 
 a crucial difference between quantum pure and mixed states. For pure states, the stationarity time $T_0$
is determined completely by the initial energy dispersion $\Delta E$. On the contrary, the value of $T_0$
for mixed states with a fixed value $\Delta E \neq 0$ can vary in wide limits, including the value $T_0=\infty$
for initial thermal states with $m\omega^2\sigma_q =\sigma_p/m= \omega\sqrt{D}>\hbar\omega/2$
(for $\omega>0$) and $\overline{q}= \overline{p}= \sigma_{qp}=0$.
The minimal value of the product (\ref{2E2T2}) for $D>\hbar^2/4$ equals $\left(4D/\hbar^2\right)^{3/2}$. It is achieved if $B\to\infty$. 
This can happen, for example, 
for highly squeezed thermal states ($\sigma_q \to 0$ or $\sigma_p \to 0$ with $D=const$, so that $\sigma_p \to \infty$
or $\sigma_q \to\infty$, respectively)  if $\overline{q}= \overline{p}=0$, 
strongly correlated mixed states ($\sigma_{qp} \to\infty$ with $D=const$ and $\overline{q}= \overline{p}=0$)
or for strongly shifted states ($\overline{q}\to\infty$ or $\overline{p}\to\infty$). 
Thus we arrive at the inequality
\be
\Delta E T_0 \ge \hbar\left(2\mu^3\right)^{-1/2}
\label{ET0G}
\ee
which holds for any Gaussian state. The equality sign takes place for free Gaussian packets with $\omega=0$.
We see that pure quantum states are 
`more fragile' than mixed states with the same energy dispersion: $T_0^{(pure)} < T_0^{(mix)}$.

Inequalities (\ref{(314)}) can be strengthened, if one takes into account the
relation $\mbox{Tr}(\hat{A}\hat{B}) \le \mbox{Tr}(\hat{A})\mbox{Tr}(\hat{B}) $,
which holds for Hermitian positively definite operators.
Introducing the `higher order purity' $\mu_n=\mbox{Tr}(\hat{\rho}^n)$, we obtain the inequalities
\be
\Delta E T_n \ge \Delta E T_0 \mu_n^{-1/2} \ge \hbar(2\mu_n)^{-1/2},
\label{(317)}
\ee
\be
\Delta E T_n \ge \Delta E T_1 \mu_{n-1}^{-1/2} \ge \hbar\mu_{n-1}^{-1/2},
\quad n\ge 2.
\label{317a}
\ee

It would be interesting to generalize inequality (\ref{ET0G}) for arbitrary (including non-Gaussian)
mixed quantum states. Probably, a simple function $\mu^{-3/2}$ can be replaced by some more complicated
dependence on the quantum purity. Such a hypothesis follows from a similar example related to the
Heisenberg--Kennard--Weyl inequality (\ref{xph2}). The product $\Delta x \Delta p$ equals $\hbar/(2\mu)$
for thermal states of a quantum oscillator, but the general `purity-dependent uncertainty relation', obtained in
\cite{183,Dod-mix}, has the form $\Delta x \Delta p \ge \Phi(\mu) \hbar/2$, where function $\Phi(\mu)$ has rather complicated
explicit expression, going asymptotically to $\Phi(\mu) \approx 8/(9\mu)$ for $\mu \ll 1$.
Perhaps, simple inequalities (\ref{(317)}) and (\ref{317a}) can be also strengthened and generalized.

\subsection{Remarks on open (dissipative) quantum systems}

It is worth noting that both the Mandelstam--Tamm inequality (\ref{(299)})  and inequalities 
(\ref{ET0G})--(\ref{317a}) are essentially based on the fact that the evolution of a
quantum system is {\em unitary\/}, being described by the linear Schr\"odinger equation or its consequence (\ref{(300)}), 
as well as on identifying the energy with the (Hermitian) Hamiltonian. For
other kinds of evolution equations, e.g., for the `standard master equation' 
\footnote{This equation was derived by several authors in 1960s \cite{Haake,Bausch,Lax}; but nowadays it is called
frequently as `Lindblad equation' after paper \cite{Lind76}.}
\be
\frac{\partial\hat\rho}{\partial t} =
\sum_{j=1}^{N}\gamma_j\left[2\hat{A}_j\hat\rho \hat{A}_j^{\dagger} - \hat{A}_j^{\dagger}\hat{A}_j\hat\rho 
-\hat\rho \hat{A}_j^{\dagger}\hat{A}_j\right]
 -\,\frac{i}{\hbar}[\hat{H}, \hat\rho]
\label{(315)}
\ee
describing open (dissipative) quantum systems, inequalities (\ref{(299)}) and 
 (\ref{ET0G})--(\ref{317a}) do not hold. This can be easily seen already in the simplest example
of the relaxation of a quantum harmonic oscillator due to the contact with a thermal bath at zero
absolute temperature. Suppose for simplicity that the initial density matrix was diagonal in the Fock basis.
Then it is known that all off-diagonal elements will remain zero with the course of time, whereas the evolution of the
diagonal elements (level population probabilities) $p_n \equiv \rho_{nn}=\langle n|\hat\rho|n\rangle$
is given by the equation [following from (\ref{(315)}) in the case of $j=1$ and $\hat{A}_1=\hat{a}$ --
the boson annihilation operator]
\footnote{
This equation was derived for the first time by Landau \cite{Land27} and generalized later by several
authors in 1960s: see, e.g., \cite{George,Scully}.}
\be
dp_n/dt = 2\gamma\left[(n+1)P_{n+1} - np_n\right], \quad n=0,1,\ldots
\label{dpn}
\ee
For the initilal Fock state $|M\rangle$ with $M \neq 0$, the initial energy dispersion $\Delta E=0$, 
while 
\[
T_0^{-2}=\sum_{n=0}^{\infty}(dp_n/dt)^2 =(2\gamma M)^2
\]
 at the initial instant $t=0$.
Consequently, $\Delta E T_0 =0$ in this case. 

It was suggested in \cite{Beretta} that the irreversible evolution can be compatible with the time--energy uncertainty
relation for some class of {\em nonlinear\/} evolution equations for the statistical operator $\hat\rho$, with a
suitable definition of the characteristic evolution time.
However, the situation is different for many nonlinear generalizations of the Schr\"odinger equation,
which were proposed in attempts to describe effects of dissipation in quantum mechanics in terms
of the wave function (see, e.g., review \cite{DoMi}). The concrete
example of equation proposed by Kostin \cite{Kostin72}
\be
\label{nlkostin}
\fl
i\hbar \frac{\partial \psi }{\partial t}= \hat{H}_{ef}\psi \equiv
 -\frac{\hbar ^2}{2m}\nabla ^2 \psi +V({\bf x}) \psi 
-i\hbar \gamma\left[ \ln \left( \frac{\psi}{\psi ^{*}}\right) -\left\langle \ln \left( \frac{\psi}{ \psi ^{*}}\right)
\right\rangle\right]\psi 
\ee
was studied in \cite{DKM-Kostin} for the harmonic oscillator potential in one dimension $V(x)=m\omega^2 x^2/2$.
Equation (\ref{nlkostin}) admits exact solutions in the form of time-dependent Gaussian wave packets. Analyzing such
solutions and calculating the time-dependent characteristic time $T_0(t)$ (\ref{(301)}), it was discovered that equality
\be
\left(\Delta {H}_{ef}\right)_t^2 T_0^2(t) =\hbar^2/2
\label{HTKost}
\ee
is fulfilled for any solution at any time instant. However, the physical meaning of the variance 
$\left(\Delta {H}_{ef}\right)^2$ is not clear, since the `effective Hamiltonian' $\hat{H}_{ef}$
depends on the wave function. On the other hand, if one uses the natural definition of `energy uncertainty' $\Delta E$
as the dispersion of the standard energy operator $-\frac{\hbar ^2}{2m}\nabla ^2  +V({\bf x})$ calculated with
respect to the time-dependent solution $\psi(x,t)$, then $\Delta E (t) T_0(t) \approx \hbar\omega/(2\sqrt{2}\gamma)$
in the over-damped case $|\gamma| \gg\omega $, so that this product  can be made as small as desired.

\section{Speed of quantum evolution}
\label{sec-speed}

The energy--time uncertainty relations are closely connected with the problem of finding the minimal time
to achieve some reference subspace in the Hilbert space, starting from a given initial state; in particular,
to reach some state orthogonal to the initial one. This subject was studied by many authors, so we give here
the most relevant results and references only. 
The first papers in this direction seem to be \cite{Anandan90,Anandan91,Vaidman92,Uhlmann92,Hubner93,Pfeifer93}.
In particular, considering the unitary evolution of pure quantum states governed by time-dependent Hamiltonians $\hat{H}(t)$,
the following inequality was obtained in \cite{Anandan90,Anandan91}:
\be
\langle \Delta E \rangle \Delta t_{\perp} \ge \pi \hbar/2.
\label{Anand}
\ee
Here $\Delta t_{\perp}$ is the minimal time interval necessary to achieve a state orthogonal to the initial one, whereas $\langle \Delta E \rangle$
is the time-averaged uncertainty in energy during this interval.
The discussions of (\ref{Anand}) and further generalizations (in particular, for mixed quantum states)
 can be found in
\cite{Hira94,Hira95,Pfeifer95,Franson,Vaidman98,Horesh98,Brody03,Kosin06,Andrews07,Brody11}. 
We bring here only one of results obtained in \cite{Pfeifer93}, generalizing  (\ref{330a}) and (\ref{330b}):
\be
\sin\left(\delta -h_t\right) \le |\langle \vf|\psi_t\rangle| \le \sin\left(\delta +h_t\right).
\label{Pfei}
\ee
Here $|\psi_t\rangle$ is the time-dependent pure quantum state originating from $|\psi_0\rangle$ at $t=0$,
$|\vf\rangle$ is some fixed quantum state, $\delta =\mbox{arcsin}|\langle \vf|\psi_0\rangle| $ and
\[
h_t=\hbar^{-1}\int_0^t \mbox{min}\left\{\Delta_{\vf}H(s), \Delta_{\psi_0}H(s)  \right\}ds,
\]
where $\Delta_{\vf}H \equiv \left[\langle\vf|\hat{H}^2|\vf\rangle - \langle\vf|\hat{H}|\vf\rangle^2 \right]^{1/2}$.

If $|\vf\rangle=|\psi_0\rangle$, then
\be
 |\langle \psi_0|\psi_t\rangle| \ge \cos\left(\hbar^{-1}\int_0^t \Delta_{\psi_0}H(s) ds\right).
\ee
If $|\vf\rangle \perp |\psi_0\rangle$, then
\be
 |\langle \vf|\psi_t\rangle| \le \sin\left(h_t\right).
\ee
This means, according to \cite{Pfeifer93}, that a state orthogonal to the initial one cannot be populated too fast,
and that a fast transition to an orthogonal state can happen if only {\em both\/} energy uncertainties,
$\Delta_{\vf}H(s)$ and $\Delta_{\psi_0}H(s)$, are large enough.

\subsection{Margolus--Levitin inequalities and their generalizations}

In the inequalities considered in the preceding sections the quantity $\Delta E$ was related somehow to the energy {\em dispersion\/},
defined by various ways. A new kind of relations containing the difference between the {\em mean energy\/} and the minimal possible value of energy 
for the given physical system was discovered by Margolus and Levitin \cite{Margol}.
Following them we assume in this subsection that $E=0$ is the minimal possible value of energy.
Then using the inequality
\be
\cos(x) \ge 1 -(2/\pi)[x +\sin(x)]
\ee
one can write the following relations for
the real and imaginary parts of nondecay amplitude $\chi(t)$ (\ref{326}):
\be
\mbox{Re}[\chi(t)] = \int_0^{\infty} \cos(Et/\hbar) P(E)dE  
\ge 1 - \frac{2\langle E\rangle}{\pi\hbar}t + \frac2{\pi}\mbox{Im}[\chi(t)].
\label{Margeq}
\ee
Let $T_{\perp}$ be the time it takes for $|\psi(0)\rangle$ to evolve into an
orthogonal state, so that $\chi(T_{\perp})=0$.  Then $\mbox{Re}[\chi(T_{\perp})]=\mbox{Im}[\chi(T_{\perp})]=0$.
Consequently
\be
\langle E\rangle T_{\perp} \ge \pi\hbar/2 .
\label{MargLev}
\ee
Applications of the quantum speed limit inequalities like (\ref{MargLev}) to the areas of quantum computation, quantum information
and quantum metrology were discussed in \cite{Lloyd00,Zwierz10,Krone10,Heger13,Poggi13}.

Generalizations of (\ref{MargLev}) were obtained in \cite{Giovan03,Batle05,LuoZhang05,Ziel06,Fu10,Yurtsever10,Frowis12,Deffner13a,Morley14}.
For example, Luo and Zhang \cite{LuoZhang05} introduced the quantity
\be
T_{\alpha}= \mbox{inf} \left\{t >0 : Q(t)=\alpha\right\}, \qquad 0\le \alpha \le 1
\ee
and proved that (here $\hbar=1$ for simplicity)
\be
T_{\alpha} \ge \pi\left(\frac{1-\sqrt{\alpha\left(1+4p^2/\pi^2\right)}}{2\langle E^p\rangle}  \right)^{1/p}
\label{LuoZhang}
\ee
for $ 0<p \le (\pi/2)\sqrt{(1/\alpha) -1}$.
In particular,
\be
T_0 \equiv T_{\perp} \ge {\pi}{\left(2\langle E^p\rangle\right)^{-1/p}}
\ee
(the quantity $T_0$ used here is different from that defined by equation (\ref{(301)}) ).
The main inequality used for the derivation of (\ref{LuoZhang}) is
\be
\cos(x) + (2p/\pi)\sin(x) \ge 1- 2(x/\pi)^p.
\ee
For $p=1$ inequality (\ref{LuoZhang}) can be easily derived from (\ref{Margeq}), if one takes into account that for the fixed value
$Q=|\chi|^2=\alpha$ one can write $\mbox{Re}(\chi)=\sqrt{\alpha}\cos(\phi)$ and $\mbox{Im}(\chi)=\sqrt{\alpha}\sin(\phi)$
with some phase $\phi$. Since the maximal  value of function $f(\phi)=\cos(\phi) - a\sin(\phi)$ is $\sqrt{1+a^2}$, one has
\[
1 - \frac{2\langle E\rangle}{\pi\hbar}t  \le \mbox{Re}[\chi(t)] - \frac2{\pi}\mbox{Im}[\chi(t)] \le
\sqrt{\alpha\left(1+4/\pi^2\right)}.
\]
Consequently
\[
2\langle E\rangle T_{\alpha} \ge \pi\hbar\left(1-\sqrt{\alpha\left(1+4/\pi^2\right)}\right).
\]

The following chain of generalized inequalities connecting the time $T_0$ with higher-order moments of the energy distribution
function was derived in \cite{Yurtsever10}:
\[
\frac{\langle (E-\langle E\rangle)^{2n}\rangle T_0^{2n}}{(2n)! \hbar^{2n}} \ge
\sum_{k=0}^{n-1} \frac{(-1)^{n-k}\langle (E-\langle E\rangle)^{2k}\rangle T_0^{2k}}
{(2k)! \hbar^{2k}}.
\]
Note however that the first inequality of this chain (for $n=1$) $T_0\Delta E \ge \hbar \sqrt{2}$
 is weaker than the M-T bound (\ref{Anand}), since $\sqrt{2} \approx 1.41 <\pi/2 \approx 1.57$.
The states that optimize inequality (\ref{MargLev}) or its generalizations (related to the problem of
time-optimal quantum evolution) were considered in 
\cite{Soder99,Carlini06,Levit09,Mostaf09,Chau10,Uzdin12,Andersson14}.

The situation is more complicated for mixed quantum states, especially in the case of non-unitary evolution
(for open quantum systems, for example)
 \cite{Andersson14,Kupf08,Jones10,Taddei13,Campo13,Deffner13,Wu15,Liu15}.
First of all, one has to generalize simple formula (\ref{defQ}), which seems a natural definition of the `non-decay probability'
for pure states. The authors of papers \cite{Kupf08,Campo13}
considered the `relative purity'
\be
Q_1(t) =\mbox{Tr}\left[\hat\rho(t)\hat\rho(0)\right]/\mbox{Tr}\left[\hat\rho^2(0)\right],
\label{Q1}
\ee
whereas the authors of \cite{Giovan03,Batle05,Jones10,Frowis12,Taddei13} used the so called Bures--Uhlmann fidelity \cite{Bures,Uhlm,Joz}
\be
Q_2(t)= \left[\mbox{Tr}\left(\sqrt{
\hat{\rho}_0^{1/2}\hat{\rho}_t \hat{\rho}_0^{1/2}}\right)  \right]^2
\label{d-buruhl}
\end{equation}
where operator $\hat{\rho}^{1/2}$ is defined as the unique
 positively semidefinite Hermitian operator
satisfying the relation $\left(\hat\rho^{1/2}\right)^2=\hat\rho$.
One of results of \cite{Campo13} was the inequality
\be
T_* \ge \frac{(1-Q_*) \mbox{Tr}\left(\hat\rho_0^2\right)}{\sqrt{\mbox{Tr}\left[\left(\hat{\cal L}^{\dagger}\hat\rho_0\right)^2\right]}},
\ee
where $\hat{\cal L}$ is the generator of evolution defined according to the equation $d\hat\rho/dt = \hat{\cal L}\hat\rho$ and $T_*$
is the time necessary to diminish the value of $Q_1$ from unity to $Q_*$. Obviously, the initial energy dispersion is not the decisive
factor in the case of non-unitary evolution. Qualitatively similar results in paper \cite{Taddei13} were expressed in less explicit form
in terms of the quantum Fisher information $F(t)=\mbox{Tr}\left[\hat\rho(t)\hat{L}^2(t)\right]$, where Hermitian operatot $\hat{L}(t)$ is
defined implicitly through the evolution equation $d\hat\rho/dt = \left(\hat{ L}\hat\rho +\hat\rho\hat{L}\right)/2$.
Counterexamples showing that there is no quantum speed limit for non-Hermitian Hamiltonians or a nonunitary evolution were given in 
\cite{Jones10,Bender07}.
Other inequalities relating the `distinguishability time' with the quantum Fisher information of $\hat\rho$ on the unitary path generated by 
$\hat{H}$ (and with $\Delta H$ for pure states) were obtained in \cite{Volkoff14}.

\subsection{Arrival and tunneling times}
\label{sec-tunn}

Another large field of research connected with ETUR is the problem of {\em arrival time\/} \cite{Allcock,Grot96,Ahar98,Delgado99,Fink99,Halliwell99,%
Baute00,Leon00}.
For example, it was shown in \cite{Ahar98} that the arrival time of a free particle to some fixed point in space cannot be deterined
with the precision better than $\Delta t_{arriv} > \hbar/\langle E \rangle$, where $\langle E \rangle$ is the mean initial kinetic energy.
Note that this relation resembles (\ref{MargLev}). This result was generalized to the case of arbitrary potentials in \cite{Baute00}. 
A comprehensive review of studies on the arrival time problem until 2000 can be found in \cite{MugaLeavens}, while 
for more recent publications one can consult e.g. \cite{Leavens02,Wlod02,Ruse02,Grubl02,Oppen02,Galapon04,Anastop06,Galapon08,Molotkov09,%
Genkin09,Year10,Anastop12,Kiukas12}.

The concept of `quantum delay time' (in scattering) was analyzed in \cite{Wigner55,Nussen02,deBianchi12}. The problem of
{\em tunneling time\/} (or traversal time) was considered in 
\cite{Nussen02,Hartman62,Hauge89,Olkhov92,Kleber94,Landauer94,Steinberg94,Steinberg95,Bracher95,Kobe01,Razavy03,Davies05,Ordon09,Choi13,%
Orlando14,Landsman15}.

Connections between the energy-time uncertainty relations  and the quantum Zeno effect
\cite{MisraSud77} were discussed in references \cite{HomeWhit86,Tambini95,Home97,Whit00,Luo02}.
On the other hand, the uncertainty relation in the non-strict form (\ref{ETh}) was used in \cite{KofKur00} to explain
a possibility of {\em anti-Zeno\/} effect.



\section{Problem of time operator}
\label{sec-oper}

Inequalities (\ref{ETh}) or (\ref{(299)}) could be derived immediately from the 
commutation relation
\be
[\hat{H}, \hat{T}] = -i\hbar,
\label{(363)}
\ee
if the time operator $\hat{T}$ existed. However, it was noticed by Pauli as far back as in 1926 
 that no Hermitian unbounded operator satisfying (\ref{(363)}) can exist for an {\em arbitrary\/} Hamiltonian $\hat{H}$ \cite{Pauli26}. 
This is connected with the specific property of energy spectrum of physical systems: since the spectrum of the time operator must
be undoubtedly continuous and unbounded, the same properties must also be
inherent in the spectrum of the Hamiltonian. However, energy spectra of the majority
of physical systems are bounded from below; in addition, they can be discrete. 
Nonetheless, many people tried to find some surrogates of the time operator for various {\em specific\/}
systems or in some restricted sense. For example, such a problem arises 
 in scattering theory, if one tries to define the duration of a collision \cite{Allcock,Smith}. 

 In the simplest case of a free particle Hamiltonian $\hat{H}_0=\hat{p}^2/(2m)$,
remembering the classical formula $x=pt/m + const$, one can define
a formal `time operator' as \cite{AharBohm61,Lippmann,Goto}
\be
\hat{T}_0 = m\left(\hat{p}^{-1}\hat{x} + \hat{x}\hat{p}^{-1}\right).
\label{(364)}
\ee
In the three-dimensional case Razavy \cite{Razavy67} considered the `time operator' in the form
\[
\hat{T}_R = \frac{m}{2}\left(\hat{r}\hat{p}_r +\hat{p}_r \hat{r}\right)\hat{H}_r^{-1},
\]
where $\hat{p}_r $ and $\hat{H}_r$ are the radial momentum and radial part of Hamiltonian, respectively.

Although the operators $\hat{H}_0$ and $\hat{T}_0$ formally
satisfy relation (\ref{(363)}), it is clear that operator (\ref{(364)}) has a lot of defects. First of all,
it contains a singular operator $\hat{p}^{-1}$.
In addition, although operator (\ref{(364)}) in the energy representation can be reduced to the form
$\hat{T}_0 = i\hbar\partial/\partial E$, nonetheless it is non-Hermitian in the space of functions 
used in physics usually. Indeed, the hermiticity condition demands the eigenstates
$\psi(E)$ to form a complete set in the semiaxis $E> 0$, to turn into zero at $E = 0$
and to be closed with respect to the operation $\partial/\partial E$. Such a set of functions does not exist,
as was pointed out in \cite{Allcock}. Therefore when dealing with operators like (\ref{(364)}) 
one should either ignore their unpleasant properties or define
the class of admissible wave functions in a special manner (see also \cite{Busch94,Miyamoto01} in this connection). 
The proposal to abandon the idea of unbounded time and to use instead some bounded time operators
satisfying (\ref{(363)}) was developed in \cite{Galapon02}.
 
Note that if the question about the existence of `time operator' is understood in a restricted
sense, e.g. as the question of finding an operator satisfying relation (\ref{(363)}) for
a given Hamiltonian, then such an operator can be found probably for
any Hamiltonian. In particular,  the prescription for how to do this
for one-dimensional systems was given in \cite{Goto81}. But the real problem consists in finding conditions under
which the formal `time operator' proves to be Hermitian (more precisely, self-conjugate).
Even more important is the possibility of physical interpretation of such an
operator as true time operator. For example, 
the problem of the energy spectrum boundedness does not exist
for the Hamiltonian $\hat{H}_F= \hat{p}^2/(2m) -F\hat{x}$ describing a particle moving in a uniform potential field. 
The spectrum of $\hat{H}_F$ is continuous and extends from $-\infty$ to $\infty$. The operator 
$\hat{T}_F = \hat{p}/F$ \cite{Razavy67,Busch94,Razavy}
is Hermitian and satisfies equation (\ref{(363)}), but what is its relationship to time? 
Moreover, $\hat{T}_F$ does not commute with the coordinate operator $\hat{x}$.
Thus, if it were the real time operator, this would mean that it is impossible
to determine simultaneously the coordinate of a particle and the time at which
this particle passes through the point with this coordinate. For these reasons
the difficulties with physical interpretation force us to treat most `time  operators' constructed thus far 
as purely mathematical artificial constructions without true physical meaning.
The worst feature of such `time operators' is that they are not universal. Instead, they should be
adjusted each time to the concrete Hamiltonian: compare operators $\hat{T}_0$ and $\hat{T}_F$.

Nonetheless attempts to construct operators resembling time in some restricted sense continue.
One direction is to extend the Hilbert space, thus removing the problem of lower bound of the
effective Hamiltonian \cite{Rosenbaum,Bauer83}. A time-like operator for the harmonic oscillator
was constructed in \cite{Garr70} and for a singular oscillator in \cite{Mikuta05}. More general cases of systems with discrete
energy spectra were considered in \cite{Pegg98,Galapon02a,Hall08}. In this connection, the entropic time--energy uncertainty relation was
introduced in \cite{Hall08}.

One may suppose that the time operator could arise naturally in the relativistic quantum mechanics:
if operators $\hat{x}_{\mu}$ exist for $\mu=1,2,3$, then operator $\hat{x}_0$ must exist as well
due to the relativistic invariance.
 Different approaches to
constructing a relativistic time operator can be found, for example, in studies
\cite{Sorkin79,Prug,Han83,Hussar}.
However, no unambiguous and
generally accepted results were obtained in this field.
 Perhaps the reason for this failure is that in
the relativistic case not only does the time operator not exist, but well-defined operators of the coordinates do not exist either. 
Thus, the time operator can apparently be introduced with the same degree of conventionality as the coordinate operator. 
For example, such a conventional operator was constructed by analogy with the known Newton--Wigner coordinate operator
\cite{NewtonWigner} in study \cite{Arshan85} (in the momentum representation for a free particle):
\be
\hat{T}_{rel} = -i\hbar\left(\frac{\partial}{\partial E} +\frac{E}{c^2 p}\frac{\partial}{\partial p}
+\frac{E}{2c^2 p^2} \right).
\label{(365)}
\ee
Then $\Delta E \Delta T \ge \hbar/2$, but operator $\hat{T}_{rel}$
 does not commute with the momentum operator:
\be
\left[\hat{T}_{rel}, \hat{\bf p}\right] = -i\hbar \frac{\hat{\bf p} E}{c^2 p^2}, \qquad
\left[\hat{T}_{rel}, \left|\hat{\bf p}\right|\right] = - \frac{i\hbar \hat{E}}{c^2 p}.
\label{(366)}
\ee
The consequence of these relations is the rigorous version of the known Landau--Peierls inequality 
\cite{LandPeierls}
\be
\Delta T \Delta p \ge \frac12 \hbar\langle v^{-1}\rangle \ge \hbar/(2c),
\label{(367)}
\ee
which relates the accuracy of the measurement of momentum to the duration of the measurement.
Other approaches were suggested or considered in \cite{Anton92,Braunstein96,Leon97,Kudaka99,Molotkov01,Wang03,Falciano10,Bauer14}.

For other studies related to the problem of time operator or estimation of time displacements (durations)
we can cite references 
\cite{Price-book,Engelmann64,Razavy69,Ekstein71,Kijowski,Helstrom74,Holevo78,Courbage80,Kobe94,Belavkin98,Kochan99,Egus00,%
Anton01,Ordon01,Brunetti02a,Smol03,%
Isidro05,Lamata05,Mielnik05,Arai05,Zimm06,Arai07,Courbage07,Wang07,Gozdz07,Torres07,Olkhov07,Arai08,Gomez08,Ivanov08,Caballar09,%
Rotter09,Brunetti10,Heger10,HegMuga10,Recami10,Bokes11,Champ11,Durt13,Torres13,Wang13,Gialam15,Bunao15}
where  different solutions were proposed or criticized.

\section{Measurements of energy and time}
\label{sec-meas}

The most interesting (and the most controversial) field of applications
of inequality (\ref{ETh}) seems to be quantum measurement theory. However, the
existence of a relation like (\ref{ETh}) in this field is still under question. Note that
this question apparently did not exist for the majority of creators of quantum theory. For example,
a number of famous thought experiments confirming relation (\ref{ETh}) were proposed by Bohr
in his discussions with Einstein during the Solvay congresses at the end of 1920s \cite{Bohr-sel}. Since that time his examples have been
widely reproduced in textbooks on quantum mechanics. However, Bohr's arguments contained a weak spot, since 
 he did not give {\em a strict and clear definition\/} for the quantities $\Delta E $ and $\Delta t$. 
Therefore all his relations are no more than estimates, so that inequality (\ref{ETh}) is valid only to within
an order of magnitude. For this reason authors of subsequent studies attached different meanings to the notion of the 
`uncertainty in energy'. For example, this quantity was defined either as the accuracy of energy measurements, or
 the difference of energy values at two time moments, or the uncertainty of this difference, 
or the uncertainty of the `increase' in the system's energy during the time of measurement, and so on. 
Moreover, it was not quite clear what is understood by the term `energy':
whether it is kinetic energy (as soon as  free particles only were discussed in the
majority of papers devoted to this subject, the corresponding energies could be nothing but kinetic energies), 
or the total energy, i.e., the Hamilton function. In addition, no hint was given to a method of calculating the
quantity $\Delta E$ for a given wave function or density matrix. The same can be
said about the quantity $\Delta t$. Different authors defined it as `the duration of the
measurement process', or `the uncertainty of the time moment at which the
change of energy of the system has taken place', or something else. Needless to say, these
quantities were not defined with mathematical accuracy. It is worth mentioning in this connection that, for example, 
such a quantity as `the duration of measurement process' can be defined quite unambiguously in the special ideal
case when the interaction between the system under study and the `measuring device' can be turned on and off abruptly at some given instants of time.
But the very existence of such abrupt interaction switching has been 
denied categorically by Fock in his papers \cite{Fock-62,Fock66} devoted to the defense of relation (\ref{ETh}). If,
however, the interaction is switched on smoothly, then a constructive definition of its duration, absent in most old papers, is needed.

For these reasons, the belief in the universal validity of relation (\ref{ETh}) for all measurement
processes has dwindled away, especially when some measurement schemes contradicting inequality (\ref{ETh}) 
were proposed. The first investigations in this direction were
performed by Aharonov and Bohm \cite{AharBohm61}, which were followed by many other studies, e.g.
 \cite{ArtKel,AharSaf,Wan80,Vorontsov81,DKM-meas83}.

  For the sake of definiteness we put the following
meaning to the quantity $\Delta E$: it is an estimate of the dispersion of energy
values before and after the measurement. In this case the question of the
validity or invalidity of inequality (\ref{ETh}) is equivalent to the question of the
validity of one of two statements:

1. In any measurement continuing for a finite interval of time any
quantum system inevitably changes its state, so that the characteristics of the
system before and after the measurement (for example, its energy) are inevitably different, 
and the dispersion of energy values before and after the
measurement satisfies inequality (\ref{ETh}), where $\Delta t$ is some characteristic 
duration of the measurement. Specifically, arbitrarily fast measurements of any
characteristic of a quantum system {\em without changing its state\/} (more precisely,
changing the state by some arbitrarily small amount) are impossible in principie.

2. The possibility of arbitrary fast `nondemolition' measurements does
not contradict the principles of quantum mechanics (i.e., situations are possible
when the product $\Delta E \Delta t$ is as close to zero as one desires).

Note that the problem consists in the {\em reproducibility\/} of the results of
the measurement, because a single measurement of energy can obviously be
performed (in principle) with an arbitrary accuracy (otherwise we would
arrive at a contradiction with the postulates of quantum mechanics, since the
energy observable corresponds to a Hermitian operator). The measurement,
as a rule, changes the state of a system, but it is not clear whether such a
situation takes place for any measurement, or whether somebody could invent
such a scheme that the measurement would be exact and instant and the stage
of the system would not be changed. Today this problem is important not
only from an academic (or abstract) standpoint, but also from a practical
(experimental) point of view, because it has an immediate relation to, for
example, constructing detectors of gravitational waves which ought to work
in the regime of the so-called quantum nondemolition measurements (QNDM)
\cite{Caves-QND,Brag80,Brag88,Voron94}.

In favor of the first statement we refer to classical papers (for example 
\cite{LandPeierls,Bohr-sel}) and the majority of textbooks on quantum mechanics.
 However, as has already been mentioned,  only some
examples have been discussed in that studies,, but no rigorous proof has been given (see e.g. \cite{Nikolic12}). On the other
hand, it would seem that the existence of a single counterexample is sufficient to refute inequality (\ref{ETh}). 
Although several such counterexamples were
given in studies  \cite{AharBohm61,ArtKel,AharSaf,Wan80,Vorontsov81,DKM-meas83,Busch90-2}, the situation is nonetheless not so simple. There is a difference of principle between the examples in studies \cite{LandPeierls,Bohr-sel}
and those in studies \cite{AharBohm61,ArtKel,AharSaf,Wan80,Vorontsov81,DKM-meas83,Busch90-2}. The authors of papers `in defense' 
of inequality (\ref{ETh}) considered various measurement schemes which could be realized
physically, if only in principle, in the real world. On the contrary, the papers like
\cite{AharBohm61,ArtKel,AharSaf,Wan80,Vorontsov81,DKM-meas83,Busch90-2} [so to speak `against' relation (\ref{ETh})] 
deal with some mathematical models, i.e., their authors, as a rule, invent various expressions for the
interaction Hamiltonians of the system under study with `measuring devices'
and investigate the properties of solutions to the corresponding equations.
However, the way to realize these Hamiltonians in real physical experiments
remains unclear. For example, a typical Hamiltonian of this type is as follows
\cite{AharBohm61,AharSaf,DKM-meas83}
\be
\hat{H} = \hat{p}_x^2/2m + \hat{p}_y^2/2M
+ \gamma(t) \hat{y} \hat{p}_x,
\label{(298)}
\ee
where a particle with mass $m$ and coordinate $x$ represents the system under
study, while a particle with mass $M \gg m$ (in order that it could be considered
quasi-classically) and coordinate $y$ represents `the measurement device'. The
interaction between the two systems is described by the term $\gamma(t)\hat{y}\hat{p}_x$, 
where $\gamma(t)$ is some known function of time disappearing
for $t \to \pm\infty$.
Fock \cite{Fock-62} severely criticized Hamiltonians like (\ref{(298)}) on the ground that they contain 
explicit functions of time. According to Fock, in this case we have to take into
account the quantum nature of fields which are created by means of such
time-dependent interactions (since, as was shown by Landau and Peierls \cite{LandPeierls}, only static
fields can be considered, generally speaking, classically, i.e., without taking
their quantum nature into account). However, it was shown in \cite{DKM-meas83} that we
can take into account the quantum nature of the `field' creating the interaction, 
believing $\gamma(t)$ to be the average value of some bosonic
field in a coherent state. Then the conclusion about the possibility of an
arbitrarily fast and nondemolition measurement remains valid. 
(Strictly speaking, certain divergent integrals arise, as occurs frequently in the quantum
field theory, due to the presence of an infinite number of field modes; but
cutting off the frequency spectrum from above by the frequency $\omega_{max} \sim (\Delta t)^{-1}$,
which seems quite reasonable and corresponds to the usual methods of quantum field theory,
we obtain corrections tending to zero as $\Delta t \to 0$.) Thus we believe that it is more important to
pay attention not so much to the function $\gamma(t)$ in formula (\ref{(298)}), but 
to the structure of the interaction Hamiltonian $\hat{H}_{int} \sim \hat{y}\hat{p}_x$:
how one could implement such an interaction in real experiments?! 

In other words, the answer to the question,
which of the two statements formulated above is correct, depends in a decisive
way on the answer to the following question: whether or not arbitrary
Hermitian interaction Hamiltonians can be realized in our physical world? If
the answer is `yes', then statement 2 is correct; if the answer is `no', then
most likely statement 1 is valid.

In summary, we will formulate our vision of the problem. It
is based on distinguishing between {\em quantum theory\/} as a fundamental science studying the world around us 
(and not yet completed) and {\em quantum mechanics\/} understood as a phenomenological and nonrelativistic model
which is (a1though logically closed and completed) suitable for describing a
limited class of phenomena: see, for example, the discussion of this difference
in \cite{Wichmann}. With such an understanding of quantum mechanics as a mathematical
scheme the second statement apparently seems correct. As to real cases, i.e.,
experiments which could be performed in the physical laboratory,
inequalities like (\ref{ETh}) must hold under the assumption (made implicitly by the founders of quantum mechanics)
 that the interaction Hamiltonians realizable in our world are those containing functions of the coordinates only (this
was clearly formulated by Landau and Peierls \cite{LandPeierls}: they mentioned that if one
could use any interaction Hamiltonian $\hat{H}(\hat{p}, \hat{x})$, then the momentum of a
particle could be measured for an arbitrarily short period of time without
any change in its velocity). Still we will again note that, first, the exact values
of the right- and left-hand sides of inequality (\ref{ETh}) for such `real' experiments have not been defined 
rigorously in many old papers, and secondly, the `derivation' of
this inequality occurs frequently inherent1y contradictory, especially in the
cases when people try to prove (\ref{ETh}) as a consequence of inequality (\ref{xph2}).
Such contradictions and ambiguities were analyzed in detail in \cite{DKM-meas83}.
For other studies on this subject we can recommend 
\cite{Helstrom76,Holevo-book,Braunstein96,BohrRos50,Mensky92,Peres93,Ahar00,Ahar02,Brunetti02,Massar05,Nicolic07,Mensky11,Mielnik13}.
Among many interesting results obtained in this area we bring two inequalities proposed by Vorontsov \cite{Vorontsov81}
(although they were not proven rigorously):
\be
\Delta E + \Delta\vep \ge \frac{\hbar}{\Delta t}, \qquad \Delta E  \Delta\vep \ge \left(\frac{\hbar}{2\Delta t}\right)^2.
\label{Vor}
\ee
Here $\Delta E$ is the error of the measurement of the system's energy during time interval $\Delta t$ and
$\Delta\vep$ is an increase of the uncertainty of the energy of the apparatus (the disturbance
of the energy of the apparatus). 

\section{Conclusion}
\label{sec-fin}

We have shown that the area of energy--time uncertainty relations continues to attract attention of many researchers until now,
and it remains alive almost 90 years after its birth. It received a `new breath' in the past quarter of century due to the actual
problems of quantum information theory and impressive progress of the experimental technique  in quantum optics and
atomic physics. It is impossible to describe various applications of the ETUR to numerous different physical phenomena
in this mini-review. Instead, we conclude with a few relevant references 
\cite{Bekenstein81,Schiffer91,Molotkov96,Bonif99,Sukhanov00,Kanai02,Anastop03,Bunge03,Bilenky08,Briggs08,Dragoman10,Denur10,Bilenky11,%
Madrid13,Rotter13,Manoukian14,Khalil15}.

\section*{Acknowledgments}

A partial support of the Brazilian agency CNPq is acknowledged.

\end{document}